%% file: main.tex
\documentclass[preprints,article,accept,pdftex,oneauthors]{Definitions/mdpi} 

\firstpage{1} 
\pubvolume{1}
\issuenum{1}
\articlenumber{0}
\pubyear{2022}
\copyrightyear{2022}
\datereceived{} 
\dateaccepted{} 
\datepublished{} 
\hreflink{https://doi.org/} 

\allowdisplaybreaks

\input{"preamble.tex"}

\Title{Reverse Sensitivity Analysis for Risk Modelling}

\TitleCitation{Reverse Sensitivity Analysis for Risk Modelling}

\Author{Silvana M. Pesenti$^{1}$\orcidA{}}

\AuthorNames{Silvana M. Pesenti}

\AuthorCitation{Pesenti, S. M.}

\address{%
$^{1}$ \quad Department of Statistical Sciences, University of Toronto; silvana.pesenti@utoronto.ca}

\corres{Correspondence: silvana.pesenti@utoronto.ca}

\abstract{
We consider the problem where a modeller conducts sensitivity analysis of a model consisting of random input factors, a corresponding random output of interest, and a baseline probability measure. The modeller seeks to understand how the model (the distribution of the input factors as well as the output) changes under a stress on the output's distribution. Specifically, for a stress on the output random variable, we derive the unique stressed distribution of the output that is closest in the Wasserstein distance to the baseline output's distribution and satisfies the stress. We further derive the stressed model, including the stressed distribution of the inputs, which can be calculated in a numerically efficient way from a set of baseline Monte Carlo samples and which is implemented in the \texttt{R} package \texttt{SWIM} on \texttt{CRAN}.\\
The proposed reverse sensitivity analysis framework is model-free and allows for stresses on the output such as $(a)$ the mean and variance, $(b)$ any distortion risk measure including the Value-at-Risk and Expected-Shortfall, and $(c)$ expected utility type constraints, thus making the reverse sensitivity analysis framework suitable for risk models.
}


\begin{document}

\section{Introduction}

Sensitivity analysis is indispensable for model building, model interpretation, and model validation, as it provides insight into the relationship between model inputs and outputs. A key tool used for sensitivity analysis are sensitivity measures, that assign to each model input a score, representing an input factor's ability to explain the variability of a model output's summary statistic; see \cite{Saltelli2008book} and \cite{Borgonovo2016EJOR} for an in-depth review. One of the most widely used output summary statistic is the variance, which gives rise to sensitivity measures, e.g., the Sobol indices, that apportion the uncertainty in the output's variance to input factors. In many applications, such as reliability management and financial and insurance risk management, however, the variance is not the output statistic of concern and instead quantile-base measures are used; indicatively see \cite{Tsanakas2016RA, Maume2018SPL, Asimit2019MF, Fissler2022sensitivity}. Furthermore, typical for financial risk management applications is that model inputs are subject to distributional uncertainty. Probabilistic (or global) sensitivity measures, however, tacitly assume that the model's distributional assumptions are correctly specified; indeed sensitivity measures based on the difference between conditional (on a model input) and unconditional densities (of the output) are termed ``common rationale'' \cite{Borgonovo2016RA}. Examples include indices, such as Borgonovo's sensitivity measures \cite{Borgonovo2007RESS}, the $f$-sensitivity index \cite{Rahman2016UQ}, and sensitivity indices based on the Cram\'er-von Mises distance \cite{Gamboa2018SIAMUQ}, we also refer to \cite{Plischke2019EJOR} for a detailed overview and to \cite{Gamboa2020ARXIV} for estimation of these sensitivity measures. Recently, \cite{Plischke2019EJOR} define sensitivity measures that depend only on the copula between input factors, whereas \cite{Pesenti2021RA} propose a sensitivity measure based on directional derivatives that take dependence between input factors into account. Estimating these sensitivities, however, may render difficult in application where joint observations are scarce, e.g., insurance portfolios, and their interpretation may be limited as dependence structures are commonly specified by expert opinions \cite{Denuit2006book}.

We consider an alternative sensitivity analysis framework proposed in \cite{Pesenti2019EJOR}
that (a) considers statistical summaries relevant to risk management, (b) applies to models subject to distributional uncertainty, thus instead of relying on correctly specified distributions from which to calculate sensitivity measures we derive alternative distributions that fulfil a specific probabilistic stress and are ``closest'' to the baseline distribution; and (c) studies reverse sensitivity measures. Differently to the framework proposed in \cite{Pesenti2019EJOR} who use the Kullback-Leibler divergence to quantify the closedness of probability measures, in this work we consider the Wasserstein distance of order two to measure the distance between distribution functions. The Wasserstein distance allows for more flexibility in the choice of stresses including survival probabilities (via quantiles) used in reliability analysis, risk measures employed in finance and insurance, and utility functions relevant for decision under ambiguity.

Central to the reverse sensitivity analysis framework is a \emph{baseline model}, the 3-tuple $(\X, g, \P)$, consisting of random input factors $\X = (X_1, \ldots, X_n)$, an aggregation function $g \colon \R^n \to \R$ mapping input factors to a univariate output $Y = g(\X)$, and a probability measure $\P$. The methodology has been termed \emph{reverse sensitivity analysis} by \cite{Pesenti2019EJOR} since it proceeds in a reverse fashion to classical sensitivity analysis where input factors are perturbed and the corresponding altered output is studied. Indeed, in the reverse sensitivity analysis proposed by \cite{Pesenti2019EJOR} a stress on the output's distribution is defined and changes in the input factors are monitored. The quintessence of the sensitivity analysis methodology is, however, not confined to stressing the output's distribution, it is also applicable to stressing an input factor and observing the changes in the model output and in the other inputs. Throughout the exposition, we focus on the reverse sensitivity analysis that proceeds via the following steps:
\begin{enumerate}[label = $\roman*)$]

    \item Specify a stress on the baseline distribution of the output;
    
    \item Derive the unique stressed distribution of the output that is closest in the Wasserstein distance and fulfils the stress;
    
    \item The stressed distribution induces a canonical Radon-Nikodym derivative $\frac{d\Q^*}{d\P}$; a change of measures from the baseline $\P$ to the stressed probability measure $\Q^*$;

    \item Calculate sensitivity measures that reflect an input factors' change in distribution from the baseline to the stressed model.

\end{enumerate}

Sensitivity testing using divergence measures -- in the spirit of the reverse sensitivity methodology -- has been studied by \cite{Cambou2017MF} using $f$-divergences on a finite probability space; by \cite{Pesenti2019EJOR} and \cite{Pesenti2020AAS} using the Kullback-Leibler divergence; and \cite{Makam2021IME} consider a discrete sample space combined with the $\chi^2$-divergence. It is however known that the set of distribution functions with finite $f$-divergence - e.g. the Kullback-Leibler and $\chi^2$ divergence - around a baseline distribution function depends on the baseline's tail-behaviour, thus the choice of $f$-divergence should be chosen dependent on the baseline distribution \cite{Kruse2019OR}. The Wasserstein distance on the contrary, automatically adapts to the baseline distribution function in that the Wasserstein distance penalises dissimilar distributional features such as different tail behaviour \cite{Bernard2020WP}. The Wasserstein distance has enjoyed numerous applications to quantify distributional uncertainty, see e.g., \cite{Blanchet2019MOR} and \cite{Bernard2020WP} for applications to financial risk management. In the context of uncertainty quantification, \cite{Moosmueller2021SIAMUQ} utilise the Wasserstein distance to elicit the (uncertain) aggregation map $g$ from the distributional knowledge of the inputs and outputs. \cite{Fort2021global} utilises the Wasserstein distance to introduce global sensitivity indices for computer codes whose output is a distribution function. In this manuscript we use the Wasserstein distance as it allows for different stresses compared to the Kullback-Leibler divergence. Indeed the Wasserstein distance allows for stresses on any distortion risk measures, while the Kullback-Leibler divergence   only allow for stresses on risk measures which are Value-at-Risk (VaR) and VaR and Expected Shortfall jointly, see \cite{Pesenti2019EJOR}.

This paper is structured as follows: In Section \ref{sec:preliminaries} we state the notation and definitions necessary for the exposition. Section \ref{sec-opt} introduces the optimisation problems and we derive the unique stressed distribution function of the output which has minimal Wasserstein distance to the baseline output's distribution and satisfies a stress. The considered stresses include constraints on risk measures, quantiles, expected utilities, and combinations thereof. In Section \ref{sec:RN} we characterise the canonical Radon-Nikodym derivative, induced by the stressed distribution function, and study how input factors' distributions change when moving from the baseline to the stressed model. An application of the reverse sensitivity analysis is demonstrated on a mixture model in Section \ref{sec:application}.

\section{Preliminaries}\label{sec:preliminaries}
Throughout we work on a measurable space $(\Omega,  \mathcal{A})$ and denote the sets of distribution functions with finite second moment by 
\begin{align*}
\begin{split}
    \M = \bigg\{\; G \colon \R \to [0,1]\; \bigg|\;
    & G  \text{ non-decreasing, right-continuous}\,, 
    \lim_{x \searrow - \infty} G(x) = 0 \,,
    \;\\
    &\lim_{x \nearrow + \infty} G(x) = 1\,,
    \text{ and }
     \left.\int x^2 \, dG(x) < + \infty\;
    \right\}\,,
\end{split}
\end{align*}
and the corresponding set of square-integrable (left-continuous) quantile functions by
\begin{equation*}
    \Breve{\M} = \left\{ \;\Ginv \in \Lp([0,1])\; \left|\; \Ginv\; \text{non-decreasing}\quad \& \quad
    \text{ left-continuous} \;\right\}\right.\,.
\end{equation*}
For any distribution function $G \in \M$, we denote its corresponding (left-continuous) quantile function by $\Ginv \in \Minv$, that is $\Ginv(u) = \inf\{\, y \in \R\, |\, G(y) \ge u\}$, $u \in [0,1]$, with the convention that $\inf \emptyset = + \infty$. We measure the discrepancy between distribution functions on the real line using the Wasserstein distance of order 2, defined as follows. 

\begin{Definition}[Wasserstein Distance]
The Wasserstein distance (of order 2) between two distribution functions $F_1$ and $F_2$ is defined as \cite{Villani2008book}
\begin{equation*}
W_2\left(F_1\,,\, F_2 \right)
    = \inf_{\pi\in\Pi(F_1,\,F_2)} \;\left\{\,\left(\int_{\R^2}|z_1-z_2|^2\,\pi(d z_1,d z_2)\right)^{\frac12}\, \right\},
\end{equation*}%
where $\Pi(F_1,F_2)$ denotes the set of all bivariate probability measures with marginal distributions $F_1$ and $F_2$, respectively. 
\end{Definition}
The Wasserstein distance is the minimal quadratic cost associated with transporting the distribution $F_1$ to $F_2$ using all possible couplings (bivariate distributions) with fixed marginals $F_1$ and $F_2$. The Wasserstein distance admits desirable properties to quantify model uncertainty such as the comparison of distributions with differing support, e.g., with the empirical distribution function. Moreover it is symmetric and forms a metric on the space of probability measures; we refer to \cite{Villani2008book} for an overview and properties of the Wasserstein distance. It is well known \cite{dall1956SNS} that for distributions on the real line, the Wasserstein distance admits the representation
\begin{equation*}
  W_2\left(F_1\,,\, F_2 \right)
   	= \left(\int_0^1 \left|\Finv_1(u) - \Finv_2(u) \right|^2 du\right)^{\frac12}\,.
\end{equation*}%

\section{Deriving the Stressed Distribution}\label{sec-opt}
Throughout this section we assume that the modeller's \emph{baseline model} is the 3-tuple $(\X, g, \P)$ consisting of a random vector of input factors $\X = (X_1, \ldots, X_n)$, an aggregation function $g \colon \R^n \to \R$ mapping input factors to a (for simplicity) univariate output $Y = g(\X)$, and a probability measure $\P$. The baseline probability measure $\P$ reflects the modeller's (statistical and expert) knowledge of the distribution of $\X $ and we denote the distribution function of the output by $F(y) = \P( Y \le y)$. The modeller then performs reverse sensitivity analysis, that is tries to understand how prespecified stresses/constraints on the output distribution $F$, e.g., an increase in jointly its mean and standard deviation or a risk measures such as the Value-at-Risk (VaR) or Expected Shortfall (ES), affects the baseline model, e.g., the joint distribution of the input factors. For this, we first define the notion of a \textit{stressed distribution}.
Specifically, for given constraints we call a solution to the  optimisation problem
\begin{equation}\label{opt:general}
    \argmin_{G \in \M}\; W_2(G, F) \qquad \text{subject to stresses/constraints on } G\,, \tag{P}
\end{equation}
a stressed distribution.

Next, we recall the concept of weighted isotonic projection which is intrinsically connected to the solution of optimisation problem \eqref{opt:general}; indeed the stressed quantile functions can be uniquely characterised via weighted isotonic projections.

\begin{Definition}[Weighted Isotonic Projection \cite{Barlow1972book}]
The weighted isotonic projection $\ell^{\uparrow_w}$ of a function $\ell \in \Lp([0,1])$ with weight function $ w \colon [0,1] \to [0, + \infty) $, $w \in \Lp([0,1])$, is its weighted projection onto the set of non-decreasing  and left-continuous functions in $\Lp([0,1])$. That is, the unique function satisfying
\begin{equation*}
\ell^{\uparrow_w}\, = \argmin_{h \in \Minv} \, \int_0^1 \left(\ell(u) - h(u) \right)^2\, w(u)\, du\,.
\end{equation*}
\end{Definition}
When the weight function is constant, i.e. $w(x) \equiv c$, $c > 0$, we write $\ell^\uparrow(\cdot) =\ell^{\uparrow_c}(\cdot)$, as in this case the isotonic projection is indeed independent of $c$. The weighted isotonic projection admits not only a graphical interpretation as the non-decreasing function that minimises the weighted $\Lp$-distance from $\ell$ but has also a discrete counterpart: the weighted isotonic regression \cite{Barlow1972book}. Numerically efficient algorithms for calculating weighted isotonic regressions are available; e.g., the \texttt{R} package \texttt{isotone} \cite{De2010JSS}.

\subsection{Risk Measure Constraints}
This section considers stresses on distortion risk measures, that is we derive the unique stressed distribution that satisfies an increase and/or decrease of distortion risk measures while minimising the Wasserstein distance to the baseline distribution $F$.

\begin{Definition}[Distortion Risk Measures]
Let $\gamma \in \Lp([0,1])$ be a square-integrable function with $\gamma \colon [0,1] \to [0,+ \infty)$ and $\int_0^1 \gamma(u)\, du = 1$. Then the distortion risk measure $\rho_\gamma$ with distortion weight function $\gamma$ is defined as
\begin{equation}\label{eq:risk-measure-def}
    \rho_\gamma (G) = \int_0^1 \Ginv(u) \gamma(u) \, du\, 
    \quad \text{for} \quad
    G\in \M\,.
\end{equation}
\end{Definition}
The above definition of distortion risk measures makes the assumption that positive realisations are undesirable (losses) while negative realisations are desirable (gains). The class of distortion risk measures includes one of the most widely used risk measures in financial risk management, the Expected Shortfall (ES) at level $\alpha \in[0,1)$ (also called Tail Value-at-Risk), with $\gamma(u) = \frac{1}{1 - \alpha}\Id_{\{u > \alpha\}}$, see e.g., \cite{Acerbi2002JBF}. The often used risk measure Value-at-Risk (VaR), while admitting a representation given in \eqref{eq:risk-measure-def}, has a corresponding weight function $\gamma$ that is not square-integrable. We derive the solution to optimisation problem \eqref{opt:general} with a VaR constraint in Section \ref{sec:VaR-const}.

\begin{Theorem}[Distortion Risk Measures]\label{thm:rm-constraints}
Let $r_k \in \R$, $\rho_{\gamma_k}$ be a distortion risk measure with weight function $\gamma_k$ and assume there exists a distribution function $\tilde{G}\in \M$ satisfying $\rho_{\gamma_k}(\tilde{G}) = r_k$ for all $k \in\{ 1, \ldots, d\}$. Then, the optimisation problem
\begin{equation}\label{opt:1}
    \argmin_{G \in \M}\; W_2(G, F) \qquad \text{subject to}\qquad \rho_{\gamma_k}(G) = r_k\, \quad k = 1, \ldots, d, 
\end{equation}
has a unique solution given by 
\begin{equation}\label{opt:1:sol}
    \Ginv^*(u) = \left(\Finv(u) + \sum_{k = 1}^d\lambda_k \gamma_k(u)\right)^\uparrow\,,
\end{equation}
where the Lagrange multipliers $\lambda_k$ are such that the constraints are fulfilled, that is $ \rho_{\gamma_k}(G^*) = r_k$ for all $k = 1, \ldots , d$. 
\end{Theorem}
We observe that the optimal quantile function is the isotonic projection of a weighted linear combination of the baseline's quantile function $\Finv$ and the distortion weight functions of the risk measures. A prominent group of risk measures is the class of coherent risk measures, that are risk measures fulfilling the properties of monotonicity, positive homogeneity, translation invariance, and sub-additivity; see \cite{Artzner1999MF} for a discussion and interpretation. It is well-known that a distortion risk measure is coherent, if and only if, its distortion weight function $\gamma(\cdot)$ is non-decreasing \cite{Kusuoka2001AME}. For the special case of a constraint on a coherent distortion risk measure that results in a larger risk measure compared to the baseline's, we obtain an analytical solution without the need to calculate an isotonic projection.

\begin{Proposition}[Coherent Distortion Risk Measure]\label{prop:coherent-rm-const}
If $\rho_{\gamma}$ is a coherent distortion risk measure and $r \ge \rho_{\gamma}(F)$, then optimisation problem \eqref{opt:1} with $d = 1$ has a unique solution given by
\begin{equation*}
    \Ginv^*(u) = \Finv(u) + \frac{r - \rho_\gamma(F)}{\int_0^1 \left(\gamma(u)\right)^2\, du} \;\gamma(u)\,.
\end{equation*}
\end{Proposition}
We illustrate the stressed distribution functions for constraints on distortion risk measures in the next example. Specifically, we look at the $\alpha$-$\beta$ risk measures which are a parametric family of distortion risk measures.
\begin{Example}[$\alpha$-$\beta$ Risk Measure]
The $\alpha$-$\beta$ risk measure, $0<\beta\le\alpha<1$, is defined by 
\begin{equation*}
    \gamma(u) = \tfrac{1}{\eta}\,\left( p\,\Id_{\{u< \beta \}} + (1-p)\,\Id_{\{u\ge\alpha\}}\right),
\end{equation*}
where $p\in[0,1]$ and $\eta=p\, \beta+(1-p)\,(1-\alpha)$ is the normalising constant. This parametric family contains several notable risk measures as special cases: for $p =0 $ we obtain $\ES_\alpha$, and for $p = 1$ the conditional lower tail expectation (LTE) at level $\beta$.
\begin{figure}[h]
\begin{center}
\includegraphics[width=0.32\textwidth]{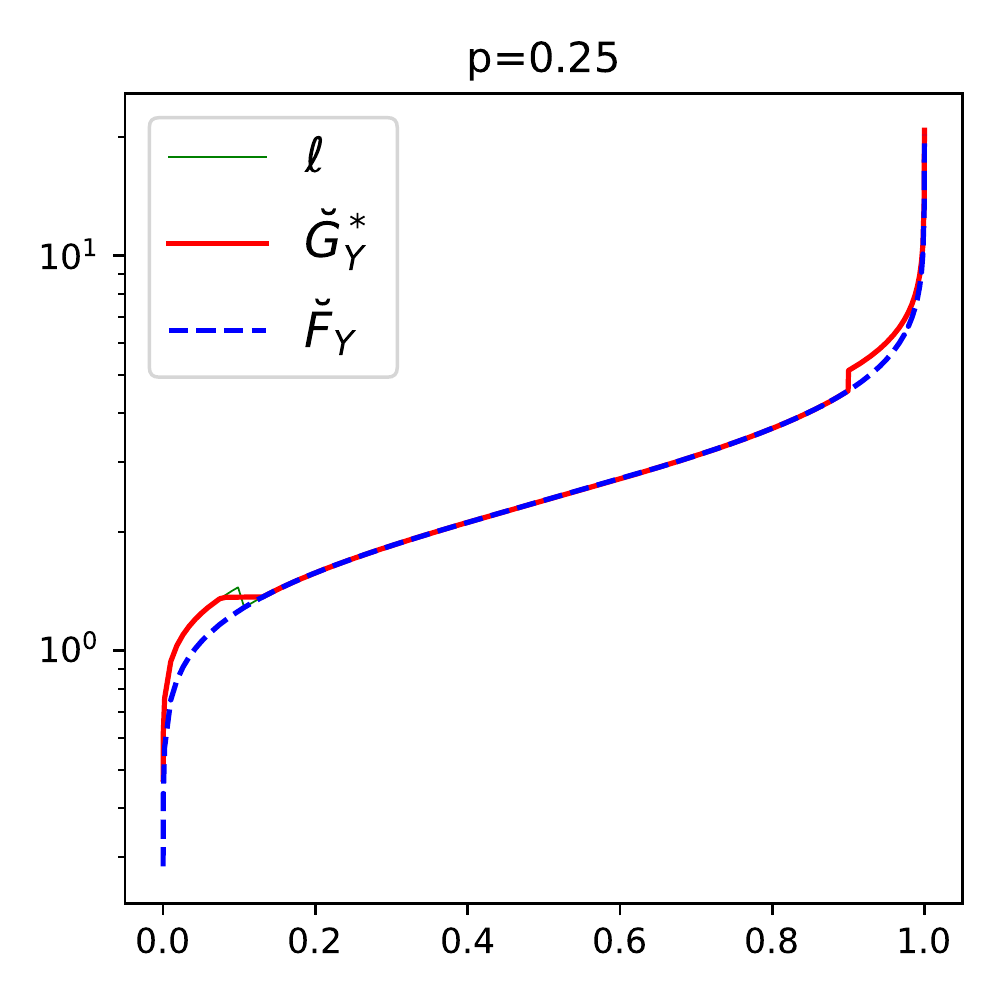}
\includegraphics[width=0.32\textwidth]{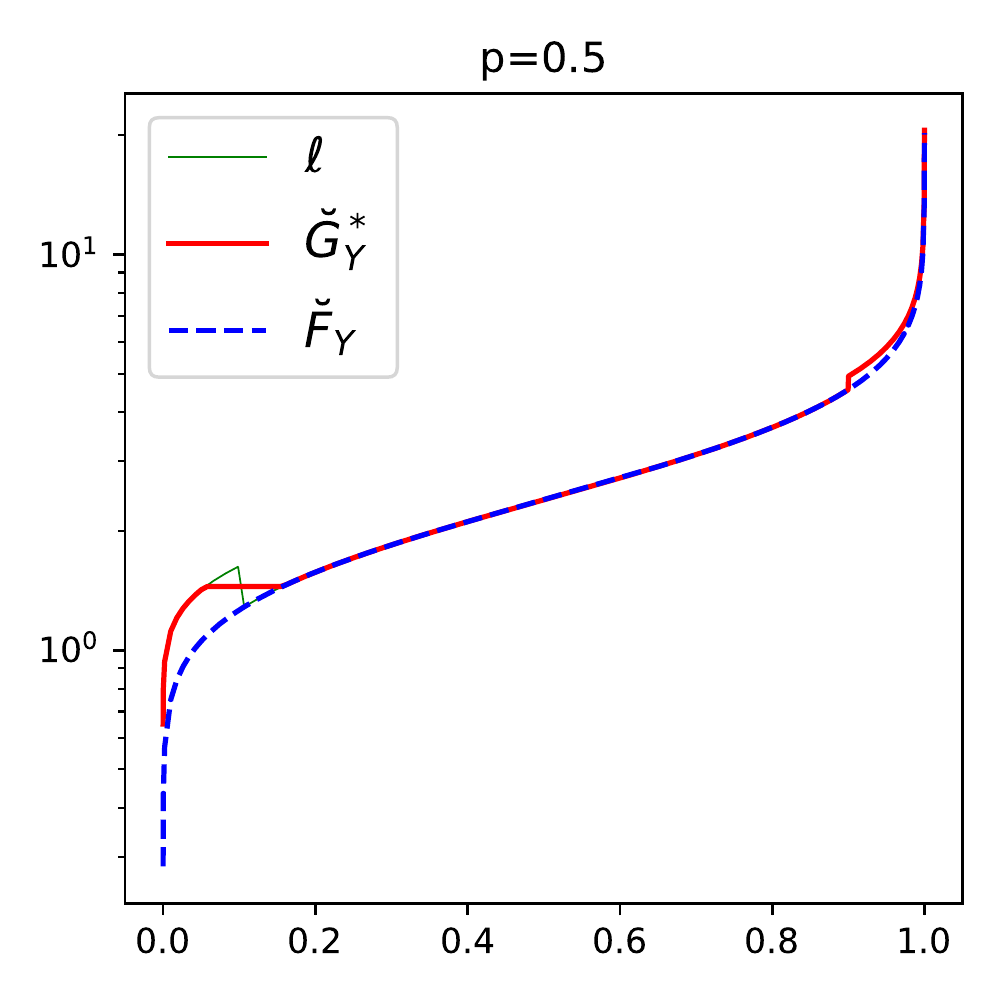}
\includegraphics[width=0.32\textwidth]{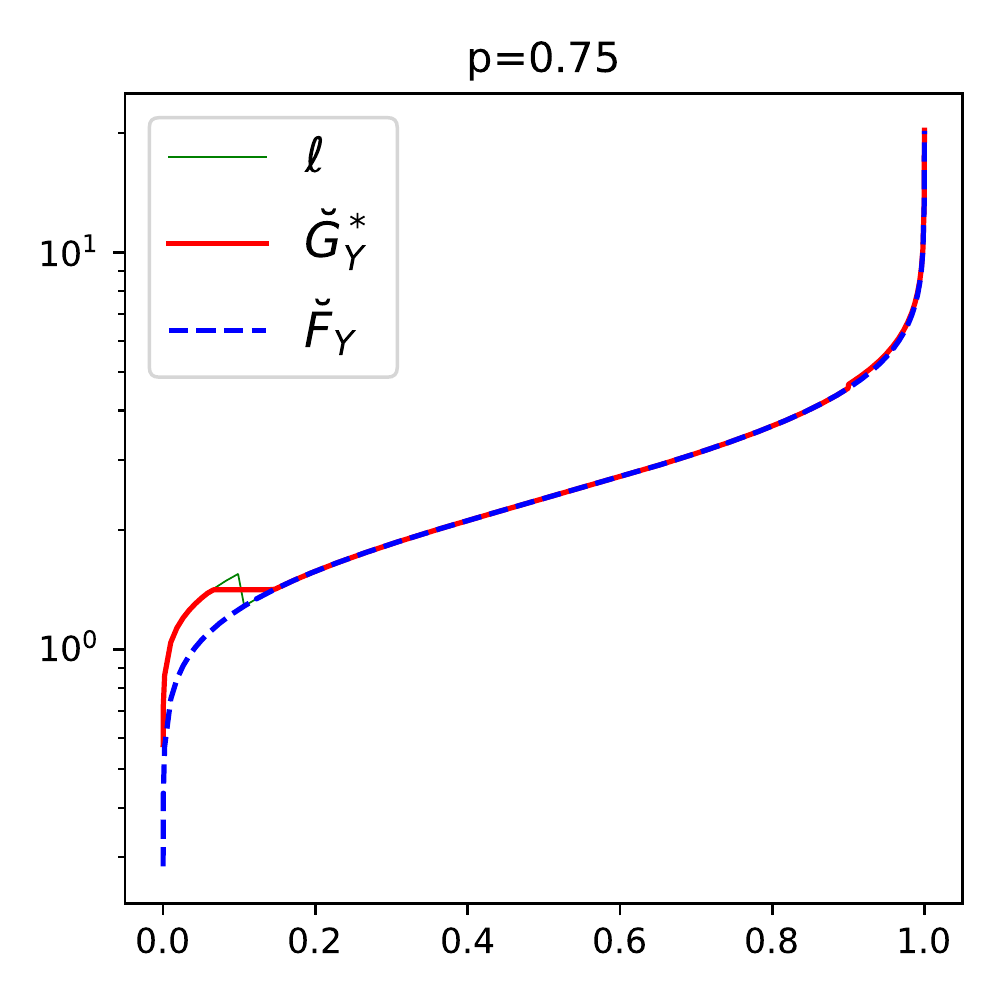}
\\[1em]
\hspace*{0.2em}
\includegraphics[width=0.31\textwidth]{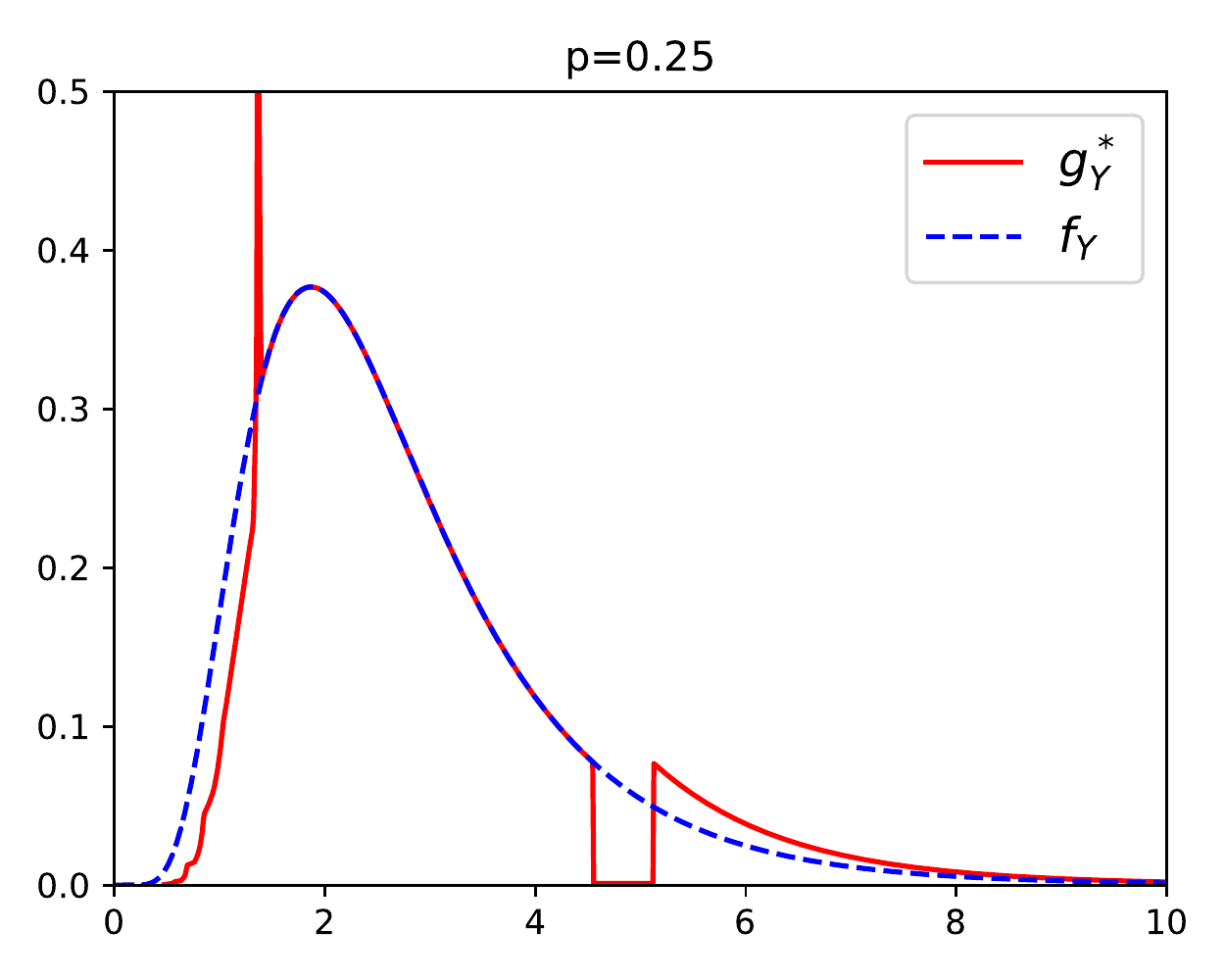}
\hspace{0.2em}
\includegraphics[width=0.31\textwidth]{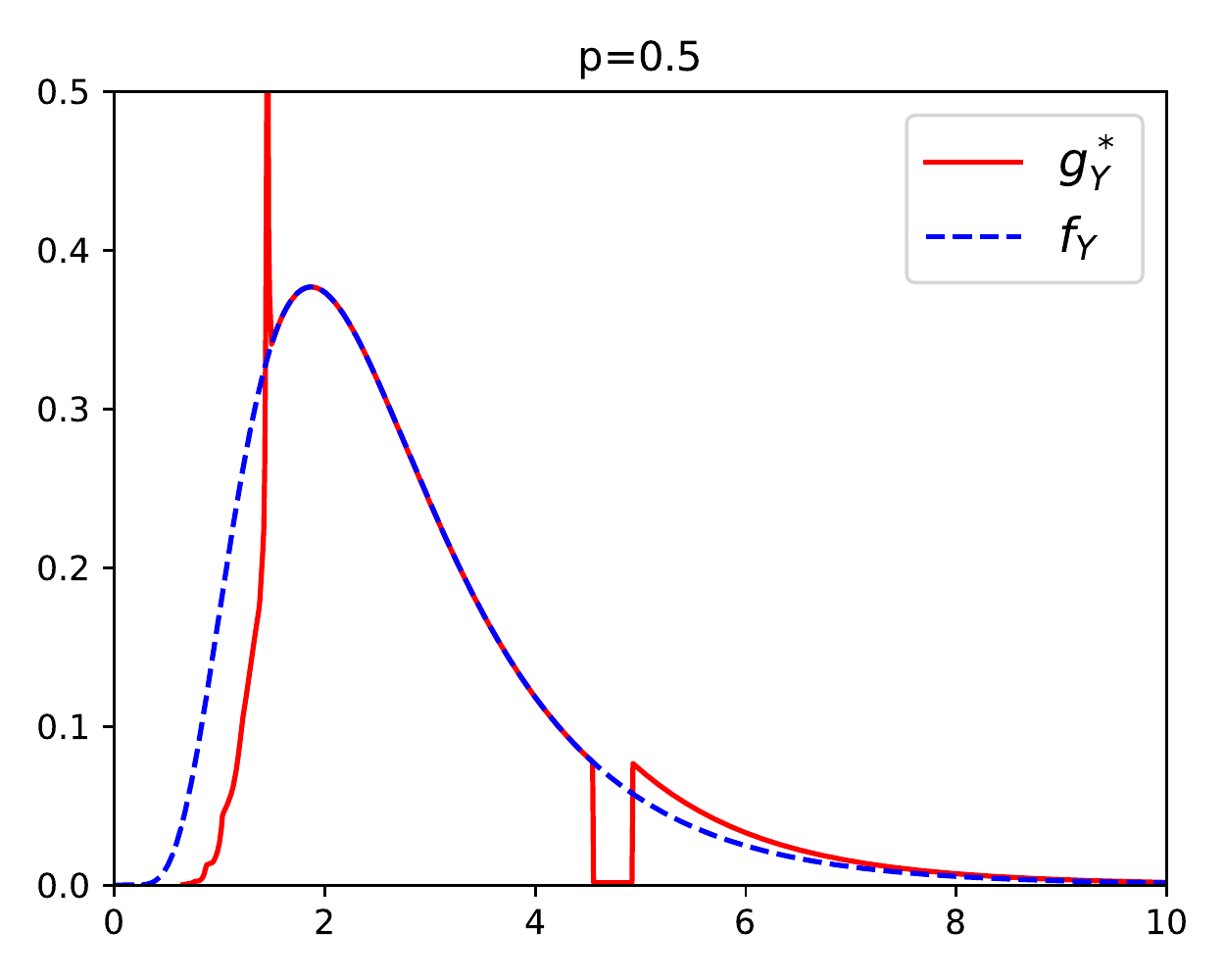}
\hspace{0.2em}
\includegraphics[width=0.31\textwidth]{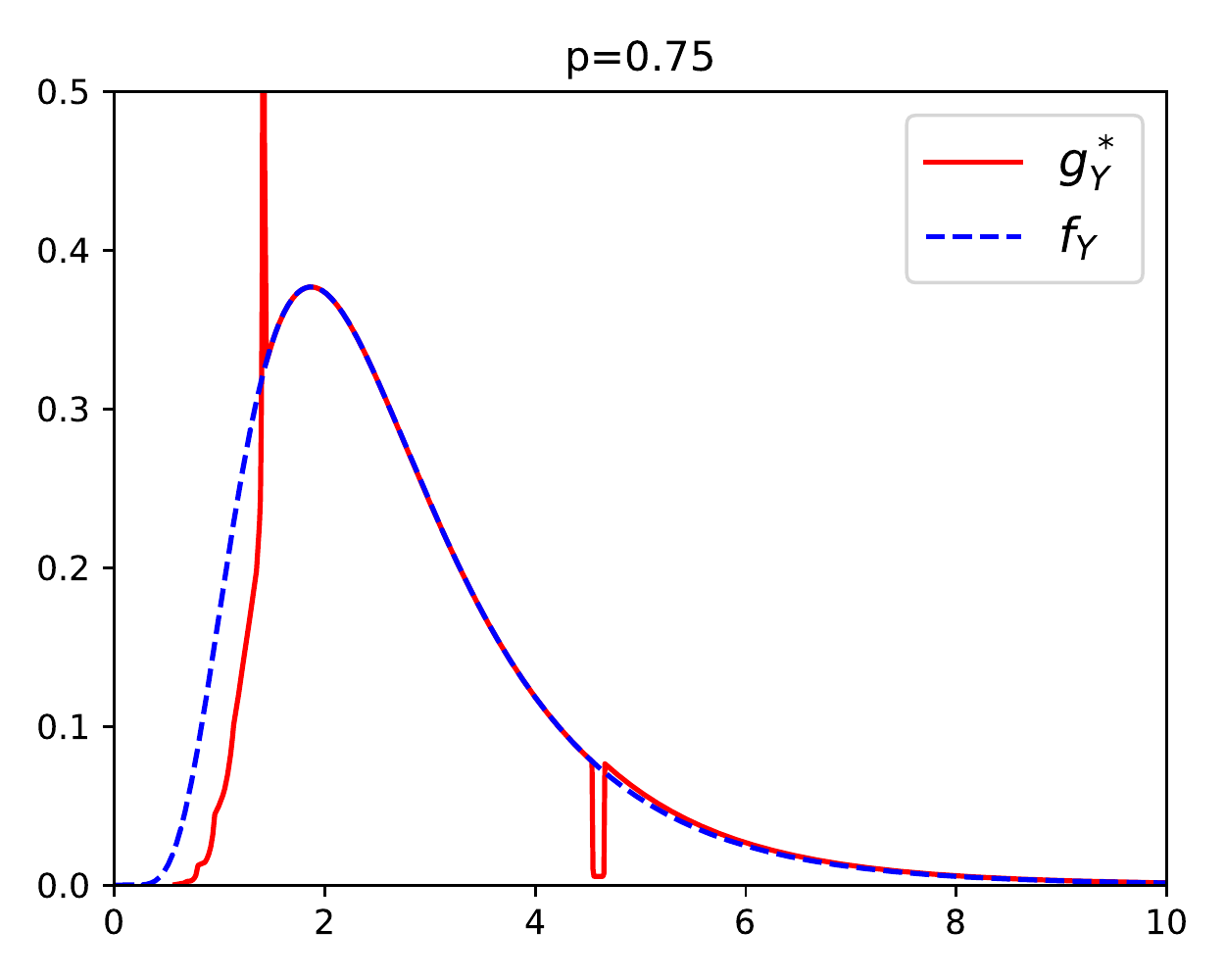}
\end{center}
\caption{Top panels: Baseline quantile function $\Finv_Y$ (blue dashed) compared to the stressed quantile function $\Ginv^*_Y$ (red solid) for a 10\% increase on the $\alpha$-$\beta$ risk measure with $\beta = 0.1$, $\alpha = 0.9$, and various values of $p$. The green line $\ell(\cdot)$ is the function, whose isotonic projection equals $\Ginv_Y(\cdot)$. Bottom panels: corresponding baseline $f_Y$ and stressed $g_Y^*$ densities.}
\label{fig:alpha-beta-various-p}
\end{figure}
Moreover, if $p <\frac{1}{2}$ $\left(\,p >\frac{1}{2}\,\right)$ the $\alpha$-$\beta$ risk measure emphasises losses (gains) relative to gains (losses). For $\alpha=\beta$ and $p< \frac{1}{2}$, the risk measure is equivalent to $\kappa\left(\ES_\alpha[Y] -\lambda\,\E[Y]\right)$, where $\kappa=\frac{(1 - 2p)\, (1 - \alpha)}{\eta}$ and $\lambda=\frac{p}{\kappa\, \eta}$.
 
Figure \ref{fig:alpha-beta-various-p} displays the baseline $\Finv_Y$ and the stressed $\Ginv^*_Y$ quantile functions of a random variable $Y$ under a 10\% increase on the $\alpha$-$\beta$ risk measure with $\beta = 0.1$, $\alpha = 0.9$, and various $p \in\{ 0.25, 0.5, 0.75\}$. The baseline distribution is chosen to be $F_Y$ is $Lognormal(\mu, \sigma^2)$ with parameters $\mu = \frac78$ and $\sigma = 0.5$. We observe in Figure \ref{fig:alpha-beta-various-p} that the stressed quantile functions $\Ginv^*_Y$ have, in all three plots, a flat part which straddles $\beta = 0.1$ and a jump at $\alpha = 0.9$. The length of the flat part is increasing with increasing $p$ while the size of the jump is decreasing with increasing $p$. This can also be seen in the stressed densities $g_Y^*$ which have, for all values of $p$, a much heavier right albeit a much lighter left tail than the density of the baseline model. Thus, under this stress, both tails of the baseline distribution are altered.
\end{Example}

\subsection{Integral Constraints}\label{sub-sec-int-constraints}
The next results are generalisations of stresses on distortion risk measures to integral constraints, and include as a special case a stress jointly on the mean, the variance, and distortion risk measures.

\begin{Theorem}[Integral]\label{thm:mult-int-constraints}
Let $h_k ,\,\tilde{h}_l\colon [0,1] \to [0,  \infty)$ be square-integrable functions and assume there exists a distribution function $\tilde{G} \in \M$ satisfying $\int_0^1 h_k(u) \Ginv(u)\, du  \le c_k$ and $\int_0^1 \tilde{h}_l(u) \left(\Ginv(u)\right)^2\, du \,  \le \tilde{c}_l$ for all $k = 1, \ldots, d$, and $l = 1, \ldots, \tilde{d}$. Then the optimisation problem
\begin{align*}
\begin{split}
    \argmin_{G \in \M}\; W_2(G, F) \qquad \text{subject to}\qquad & \int_0^1 h_k(u) \Ginv(u)\, du \,  \le c_k\,, \quad k = 1, \ldots, d, \\
    &\int_0^1 \tilde{h}_l(u) \left(\Ginv(u)\right)^2\, du \,  \le \tilde{c}_l\,, \quad  l = 1, \ldots, \tilde{d},
\end{split}
\end{align*}
has a unique solution given by 
\begin{equation*}
    \Ginv^*(u) = 
    \left(\frac{1}{\tilde{\Lambda}(u)}\left(\Finv(u) + \sum_{k = 1}^d\lambda_k h_k(u)\right) \right)^{\uparrow_{\tilde{\Lambda}}}\,,
\end{equation*}
where $\tilde{\Lambda}(u) = 1 +\sum_{k = 1}^{\tilde{d}} \tilde{\lambda}_k\tilde{h}_k(u)$ and the Lagrange multipliers $ \lambda_1,\ldots, \lambda_d$ and $ \tilde{\lambda}_1,\ldots, \tilde{\lambda}_d$ are non-negative and such that the constraints are fulfilled.
\end{Theorem}

A combination of the above theorems provides stresses jointly on the mean, the variance, and on multiple distortion risk measures. 
\begin{Proposition}[Mean, Variance, and Risk Measures]\label{prop:moment-risk-measures}
Let $m^\prime\in \R$, $\sigma^\prime >0$, $r_k \in \R$, and distortion risk measures $\rho_{\gamma_k}$, $k = 1, \ldots, d$. Assume there exists a distribution function $\tilde{G}\in\M$ with mean $m^\prime$, standard deviation $\sigma^\prime$, and which satisfies $\rho_{\gamma_k}(\tilde{G}) = r_k$, for all $k = 1, \ldots, d$.
Then the optimisation problem
\begin{align*}
\begin{split}
    \argmin_{G \in \M}\; W_2(G, F) \qquad \text{subject to}\qquad 
        & \int x \, dG(x) = m^\prime,\\
        & \int (x - m^\prime)^2\, dG(x) = \left(\sigma^\prime\right)^2 \quad \text{and}\\
        & \rho_{\gamma_k}(G) = r_k,\quad k = 1, \ldots, d, 
\end{split}
\end{align*}
has a unique solution given by 
\begin{equation*}
    \Ginv^*(u) = \left(\frac{1}{1 + \lambda_2}\left(\Finv(u) + \lambda_1 + \lambda_2 m^\prime
    + \sum_{k = 1}^{d} \lambda_{k+2} \,\gamma_k(u)\right)\right)^\uparrow\,,
\end{equation*}
and the Lagrange multipliers $\lambda_1, \ldots, \lambda_{d+2}$ with $\lambda_2 \neq -1$ are such that the constraints are fulfilled.
\end{Proposition}

\begin{Example}[Mean, Variance, \& ES]
Here, we illustrate Proposition \ref{prop:moment-risk-measures} with the ES risk measure and three different stresses. The top panels of Figure \ref{fig:mean_sd-ES-Ginv} display the baseline quantile function $\Finv_Y$ and the stressed quantile function $\Ginv^*_Y$ of $Y$, where the baseline distribution $F_Y$ of $Y$ is again $Lognormal(\mu, \sigma^2)$ with parameters $\mu = \frac78$ and $\sigma = 0.5$. The bottom panels display the corresponding baseline and stressed densities. The left panels correspond to a stress, where, under the stressed model, the $\ES_{0.95}$ and the mean are kept fixed at their corresponding values under the baseline model, while the standard deviation is increased by 20\%. We observe, both in the quantile and density plot, that the stressed distribution is more spread out indicating a larger variance. Furthermore, at $y \approx 5.77$ the stressed density $g^*_Y(y)$ drops to ensure that $\ES_{0.95}(G_Y^*) = \ES_{0.95}(F_Y)$. This drop is due to the fact that a stress composed of a 20\% increase in the standard deviation while fixing the mean (i.e., without a constraint on $\ES$) results in an ES that is larger compared to the baseline's. Indeed, under this alternative stress (without a constraint on ES) we obtain that $\ES_{0.95}(G_Y^*) \approx 7.70$ compared to $\ES_{0.95}(F_Y) \approx 6.87$. 
\begin{figure}[t]
\begin{center}
\includegraphics[width=0.32\textwidth]{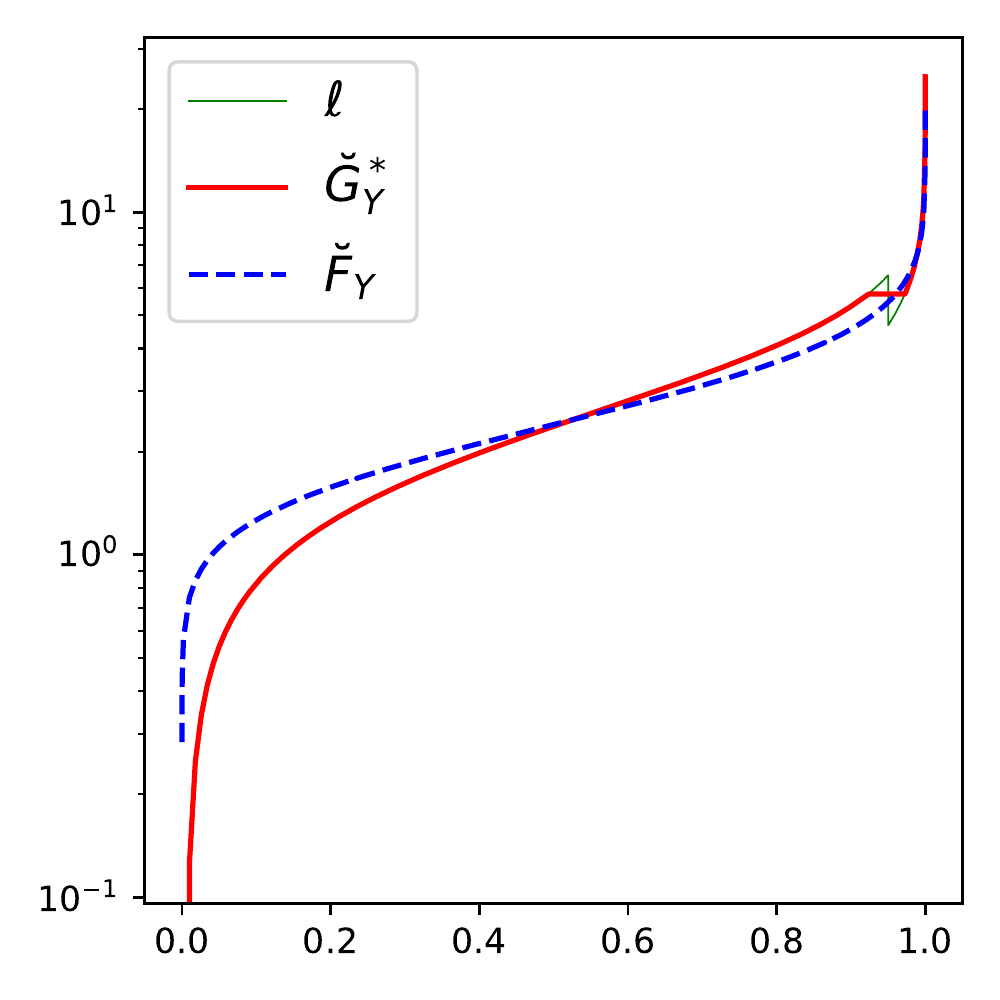}
\includegraphics[width=0.32\textwidth]{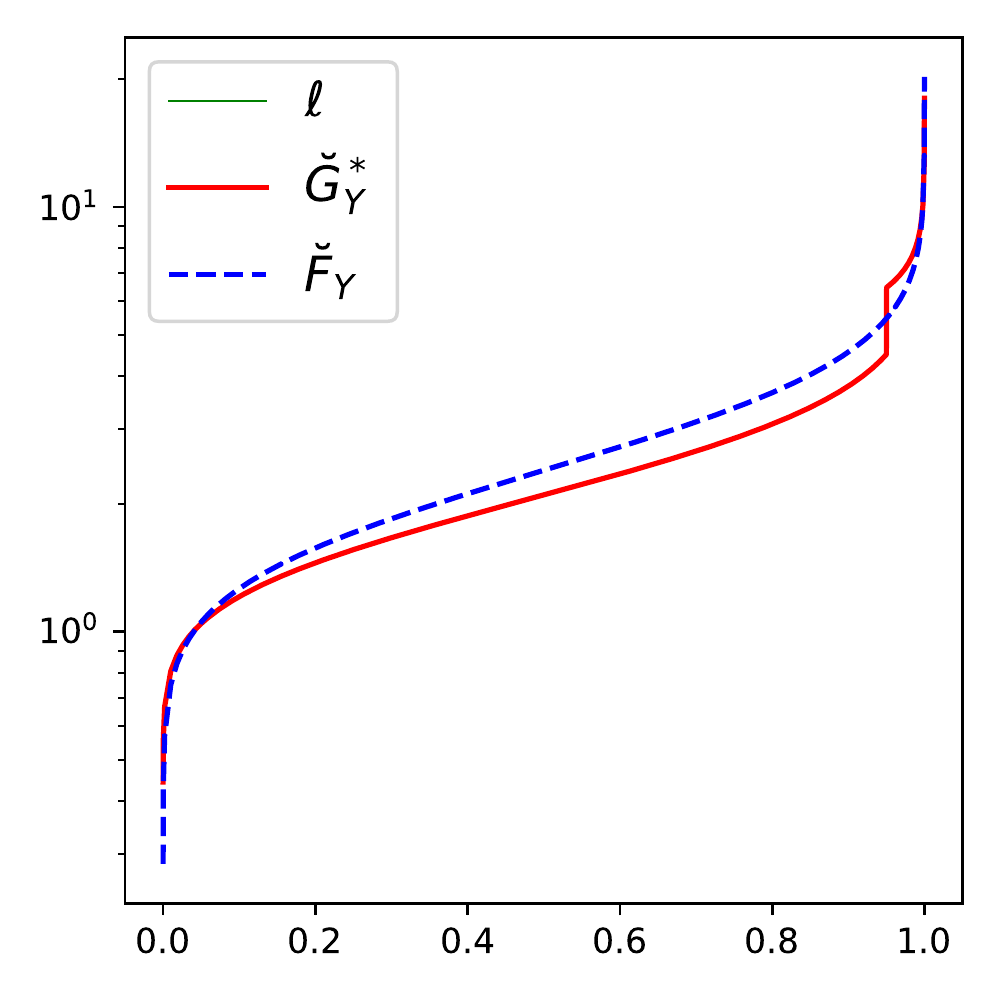}
\includegraphics[width=0.32\textwidth]{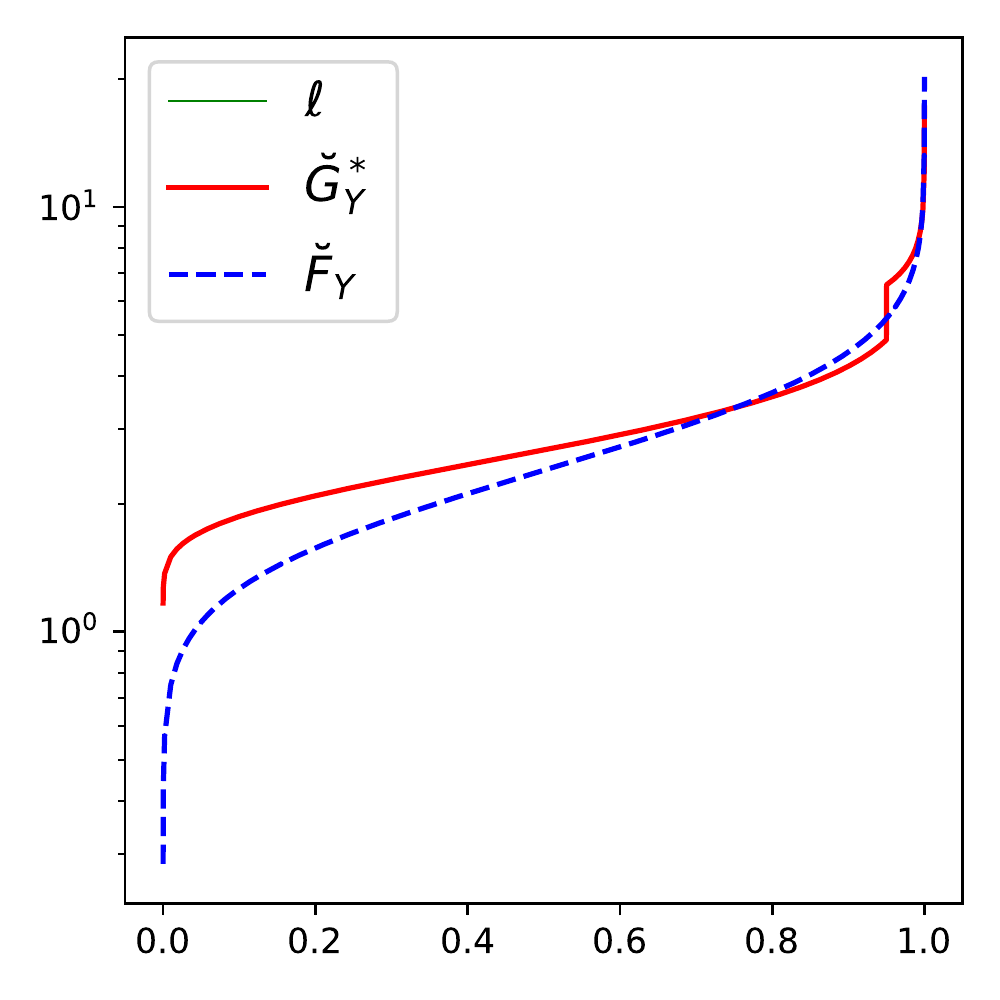}\\[1em]
\hspace*{0.2em}
\includegraphics[width=0.31\textwidth]{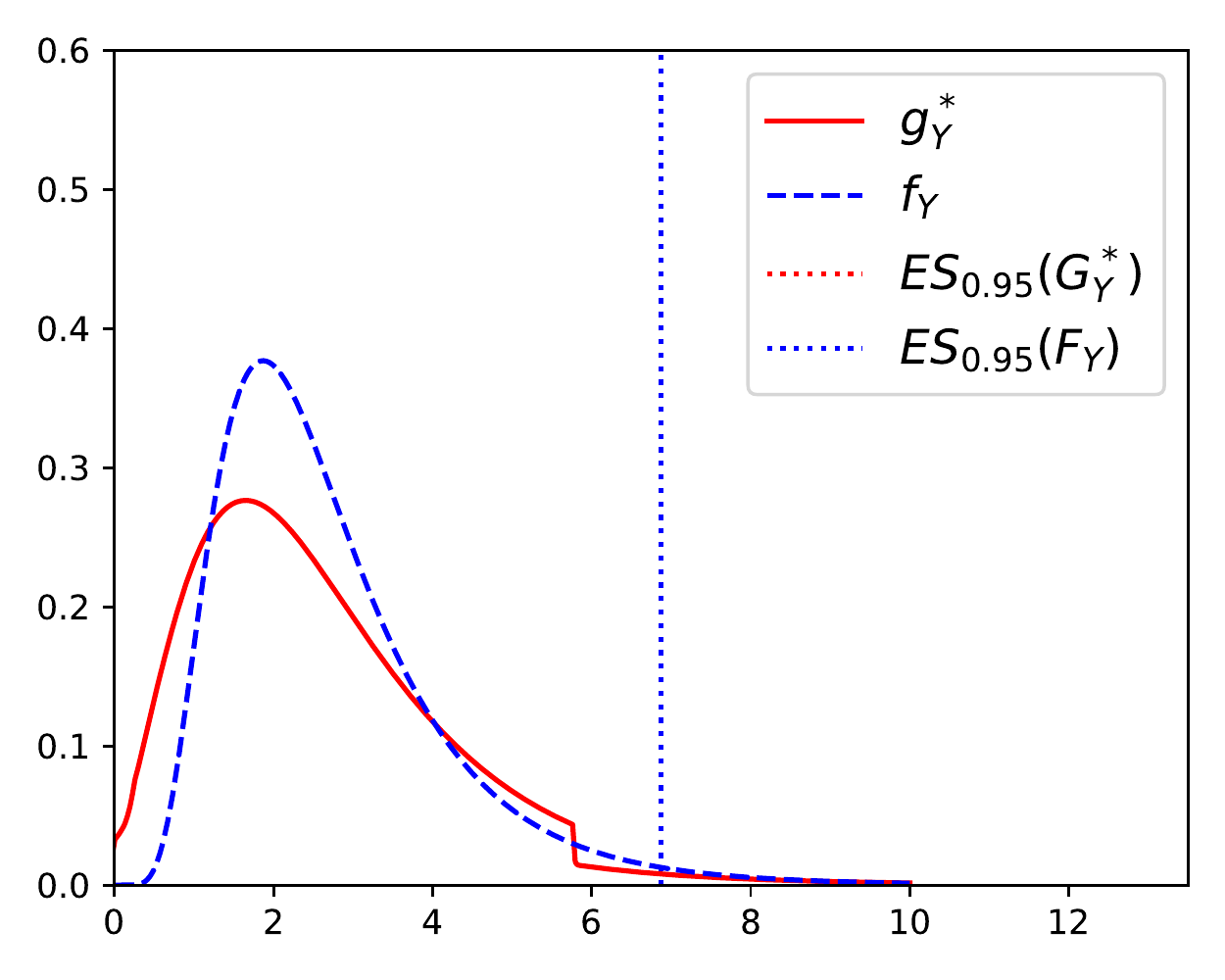}
\hspace{0.3em}
\includegraphics[width=0.31\textwidth]{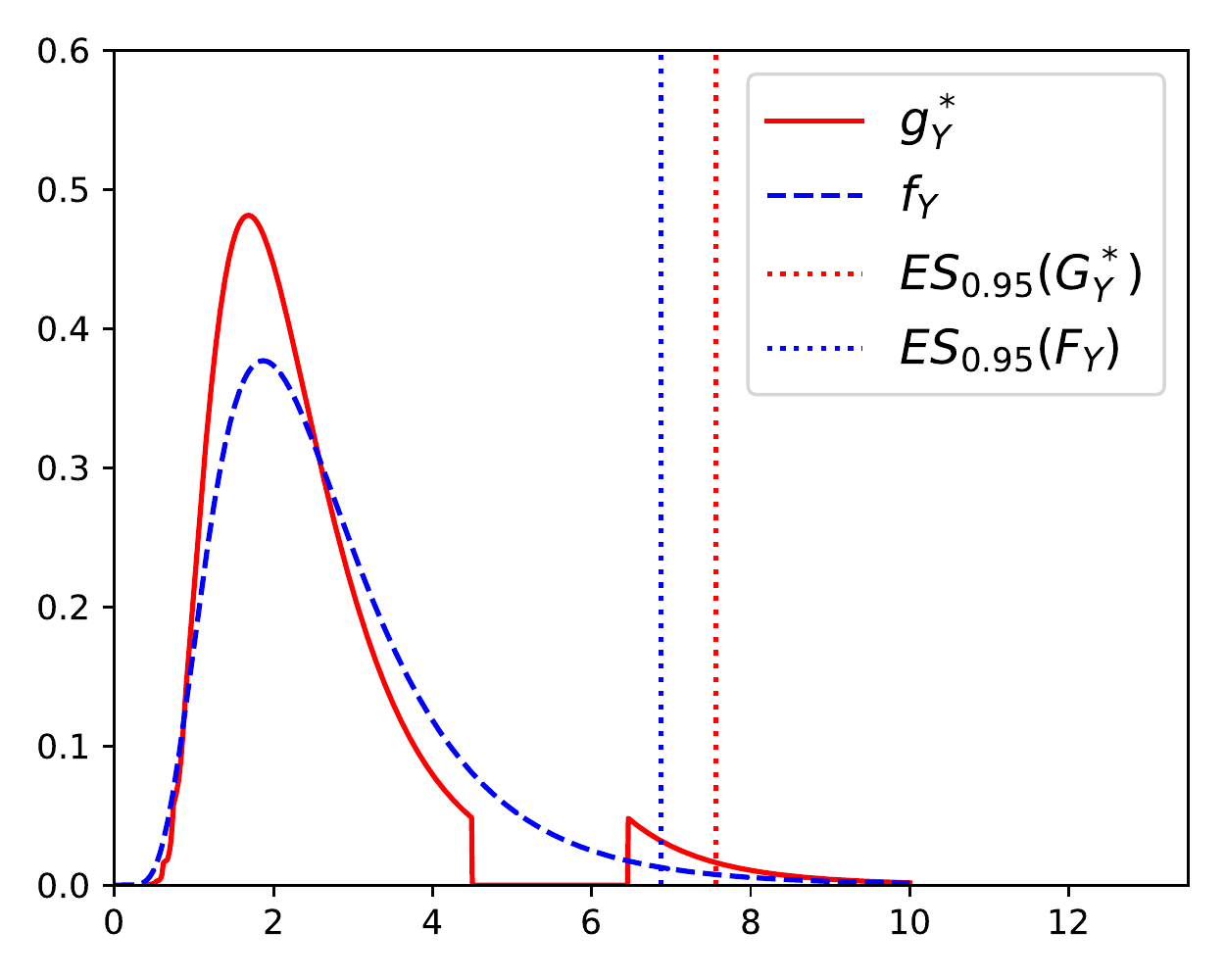}
\hspace{0.3em}
\includegraphics[width=0.31\textwidth]{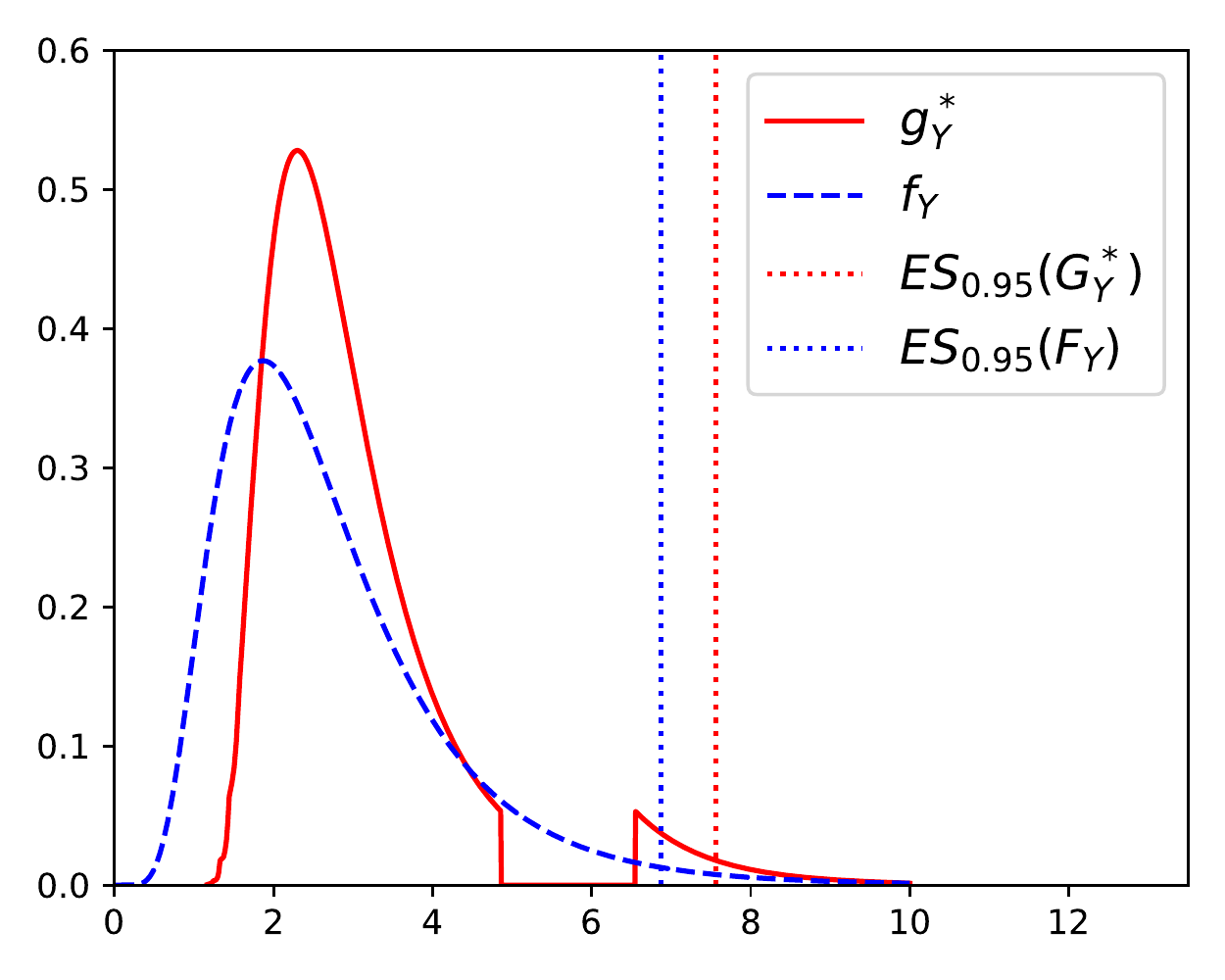}
\end{center}
\caption{Top: Baseline quantile function $\Finv_Y$ compared to the stressed quantile function $\Ginv_Y^*$. Bottom: corresponding baseline $f_Y$ and stressed $g_Y^*$ densities.
Left: $\ES_{0.95}$ and the mean being fixed and a 20\% increase in the standard deviation. 
Middle: 10\% increase in $\ES_{0.95}$, 10\% decrease in the mean, and fixed standard deviation. 
Right: 10\% increase in $\ES_{0.95}$, 10\% increase in the mean, and 10\% decrease in standard deviation. 
}
\label{fig:mean_sd-ES-Ginv}
\end{figure}
The middle panels correspond to a 10\% increase in $\ES_{0.95}$ and a 10\% decrease in the mean, while keeping the standard deviation fixed at its value under the baseline model. The density plot clearly indicates a general shift of the stressed density to the left, stemming from the decrease in the mean, and a single trough which is induced by the increase in ES. The right panels correspond to a 10\% increase in $\ES_{0.95}$, a  10\% increase in the mean, and a  10\% decrease in the standard deviation. The stressed density still has the trough from the increase in ES, however, the density is less spread out (reduction in the standard deviation) and generally shifted to the right (increase in the mean).
\end{Example}

\subsection{Value-at-Risk Constraints}\label{sec:VaR-const}
In this section we study stresses on the risk measure Value-at-Risk (VaR). The VaR at level $\alpha \in (0,1)$ of a distribution function $G \in \M$ is defined as its left-continuous quantile function evaluated at $\alpha$, that is
\begin{equation*}
    \VaR_\alpha(G) =  \Ginv(\alpha)\,.
\end{equation*}
We further define the right-continuous $\VaR^+$, that is the right-continuous quantile function of $G\in \M$ evaluated at $\alpha$, by
\begin{equation*}
    \VaR^+_\alpha(G)
    = \Ginv^+(\alpha) 
    = \inf \left.\left\{ y \in \R \;\right|\; F(y) > \alpha\right\}\,.
\end{equation*}

\begin{Theorem}[$\VaR$]\label{thm:VaR}
Let $q \in \R$ and consider the optimisation problem 
\begin{subequations}
\begin{align*}
    \argmin_{G \in \M}\; W_2(G, F) \qquad \text{subject to}\qquad  & a) \quad \VaR_\alpha(G) = q \quad \text{or} \\
     & b) \quad \VaR^+_\alpha(G) = q\,,
\end{align*}
\end{subequations}
and define $\alpha_F $ such that $\VaR_{\alpha_F}(F) = q$. Then, the following holds
\begin{enumerate}[label = $\roman*)$]
    \item under constraint a), if $q \le \VaR_\alpha (F)$, then the unique solution is given by 
\begin{equation*}
    \Ginv^*(u) = \Finv(u)  + \big(q - \Finv(u)\big)\Id_{\left\{u \in \left(\alpha_F, \alpha\right]\right\}}\,;
\end{equation*}
if $q > \VaR_\alpha (F)$, then there does not exist a solution.

    \item under constraint b), if $q \ge \VaR_\alpha^+ (F)$, then the unique solution is given by 
\begin{equation*}
    \Ginv^*(u) = \Finv(u)  + \big(q - \Finv(u)\big)\Id_{\left\{u \in \left(\alpha, \alpha_F\right]\right\}}\,;
\end{equation*}
if $q < \VaR^+_\alpha (F)$, then there does not exist a solution.
\end{enumerate}
\end{Theorem}

The above theorem states that if the optimal quantile function exists it is either the baseline quantile function $\Finv$ or constant equal to $q$. Moreover, the stressed quantile function (if it exists) jumps at $\alpha$ which implies that the existence of a solution hinges on the careful choice of the stress. For a stress on $\VaR$ (constraint a)) for example, a solution exists if and only if the constraint satisfies $q \le \VaR_\alpha(F)$; a decrease in the $\VaR_\alpha$ from the baseline to the stressed model. The reason for the non-existence of a solution when stressing VaR upwards is that the unique increasing function that minimises the Wasserstein distance and satisfies the constraint is not left-continuous and thus not a quantile function. 

Alternatively to stressing VaR or $\VaR^+$, and in particularly in the case when a desired stressed solution does not exist, one may stress instead the distortion risk measure Range-Value-at-Risk (RVaR) \cite{Cont2010QF}. The RVaR at levels $0 \le \alpha < \beta \le 1$ is defined by
\begin{equation*}
    \RVaR_{\alpha, \beta} (G)
    = 
    \frac{1}{\beta - \alpha}\int_\alpha^\beta \Ginv(u)\, du\,,
    \quad \text{for} \quad 
    G \in \M\,,
\end{equation*}
and belongs to the class of distortion risk measures. The $\RVaR$ attains as limiting cases the VaR and $\VaR^+$. Indeed, for any $G\in \M$ it holds
\begin{equation*}
    \text{VaR}_\alpha(G) = \lim_{\alpha^\prime \nearrow \alpha}\text{RVaR}_{\alpha^\prime, \alpha}(G)\, \quad \text{and} \quad
    \text{VaR}^+_\alpha(G) = \lim_{\beta\searrow \alpha}\text{RVaR}_{\alpha, \beta}(G)\, .
\end{equation*}
The solution to stressing $\RVaR$ is provided in Theorem \ref{thm:rm-constraints}.

\subsection{Expected Utility Constraint}
This section considers the change from the baseline to the stressed distribution under an increase of an expected utility constraint. In the context of utility maximisation, the next theorem provides a way to construct stressed models with a larger utility compared to the baseline. 

\begin{Theorem}[Expected Utility and Risk Measures]\label{thm:exp-utility-risk-measure}
Let $u\colon \R \to \R$ be a differentiable concave utility function, $r_k \in \R$, and $\rho_{\gamma_k}$ be distortion risk measures, for $k = 1, \ldots, d$. Assume there exists a distribution function $\tilde{G}$ satisfying $\int_\R u(x) \, dG(x) \ge c$ and  $\rho_{\gamma_k}(G) = r_k$ for all $k = 1, \ldots, d$. 
Then the optimisation problem
\begin{align*}
    \argmin_{G \in \M}\; W_2(G, F) \qquad \text{subject to}\qquad 
    & \int_\R u(x) \, dG(x) \ge c\;\quad \& \quad\;
    \rho_{\gamma_k}(G) = r_k, \quad k = 1, \ldots, d\,
\end{align*}
has a unique solution given by 
\begin{equation}\label{eq:utility-solution}
    \Ginv^*(u) = \breve{\nu}_{\lambda_1}\left(\left(\Finv(u)  + \sum_{k = 1}^{d}\lambda_{k+1} \gamma_k(v)\right)^
    \uparrow\right)\,,
\end{equation}
where $\breve{\nu}_{\lambda_1}$ is the left-inverse of $\nu_{\lambda_1}(x ) = x  - \lambda_1 \,u^\prime(x)$, and $\lambda_1 \ge 0$,  $(\lambda_2, \ldots, \lambda_{d+1}) \in \R^d$ are such that the constraints are fulfilled. 
\end{Theorem}

The utility function in Theorem \ref{thm:exp-utility-risk-measure} need not be monotone, indeed the theorem applies to any differentiable concave function, without the need of an utility interpretation. Moreover, Theorem \ref{thm:exp-utility-risk-measure} also applies to differentiable convex (disutility) functions $\tilde{u}$ and constraint $\int_\R \tilde{u}(x) \, dG(x) \le c$; a situation of interest in insurance premium calculations. In this case the solution is given by \eqref{eq:utility-solution} with $u(x) = - \tilde{u}(x)$.

\begin{Example}[HARA Utility \& ES]\label{ex: hara-ES}
The Hyperbolic absolute risk aversion (HARA) utility function is defined by
\begin{equation*}
    u(x) = \frac{1 - \eta}{\eta}\left( \frac{a x}{1 - \eta} + b\right)^\eta\,,
\end{equation*}
with parameters $a>0$, $\frac{a x}{1 - \eta} + b>0$, and where $\eta \le 1$ guarantees concavity. 
\begin{figure}[h]
\centering
\includegraphics[width=0.32\textwidth]{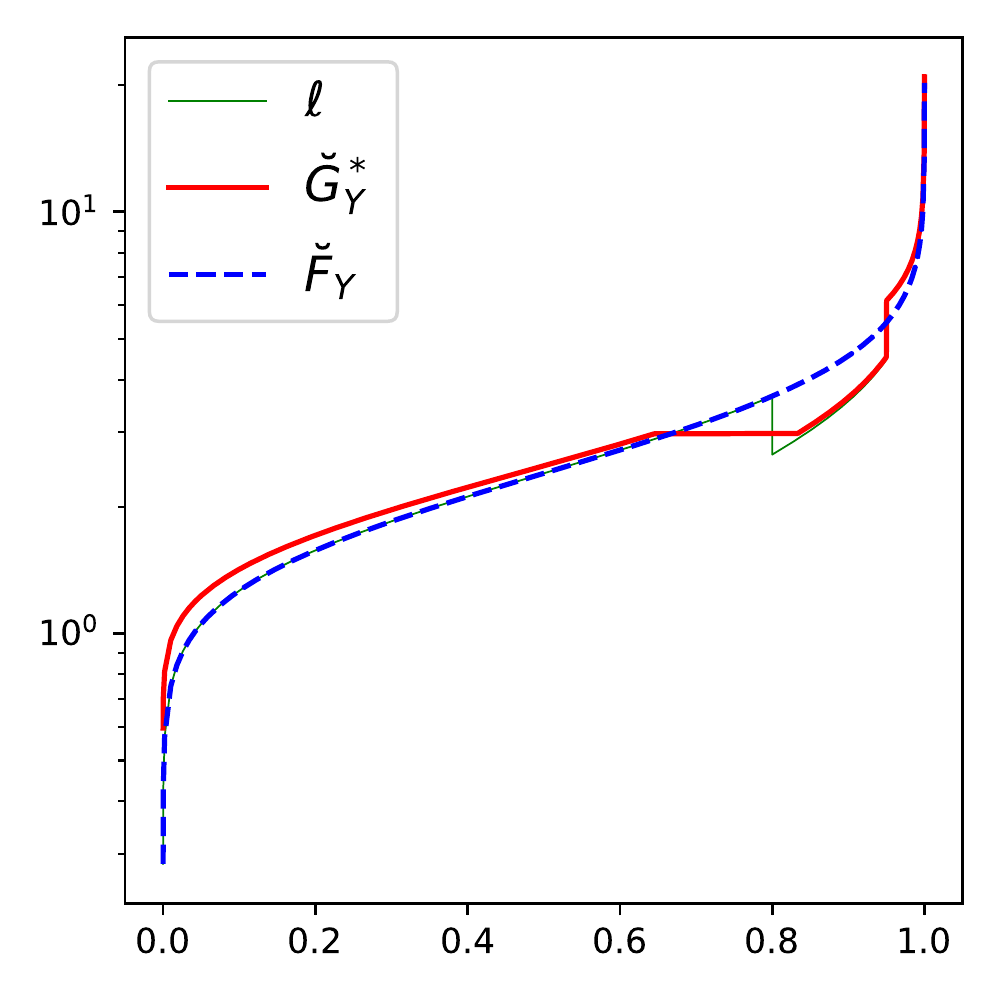}
\includegraphics[width=0.32\textwidth]{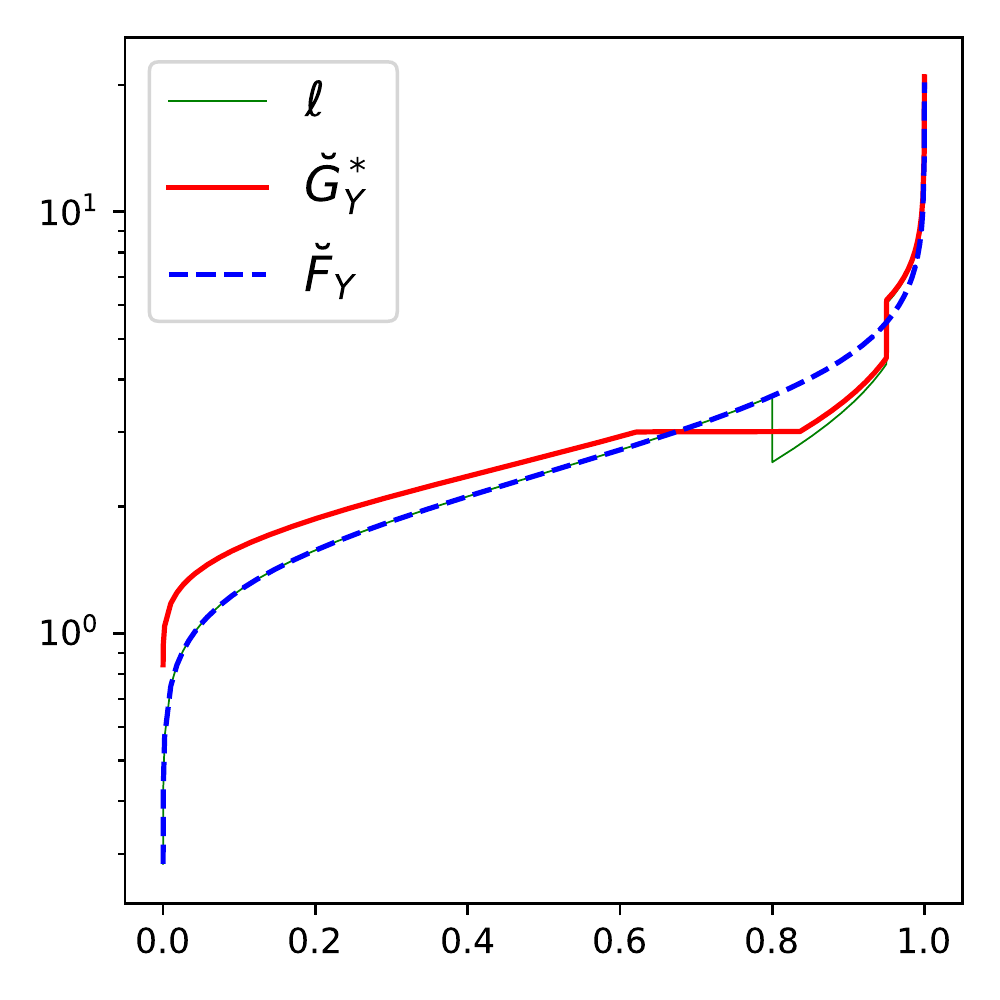}
\includegraphics[width=0.32\textwidth]{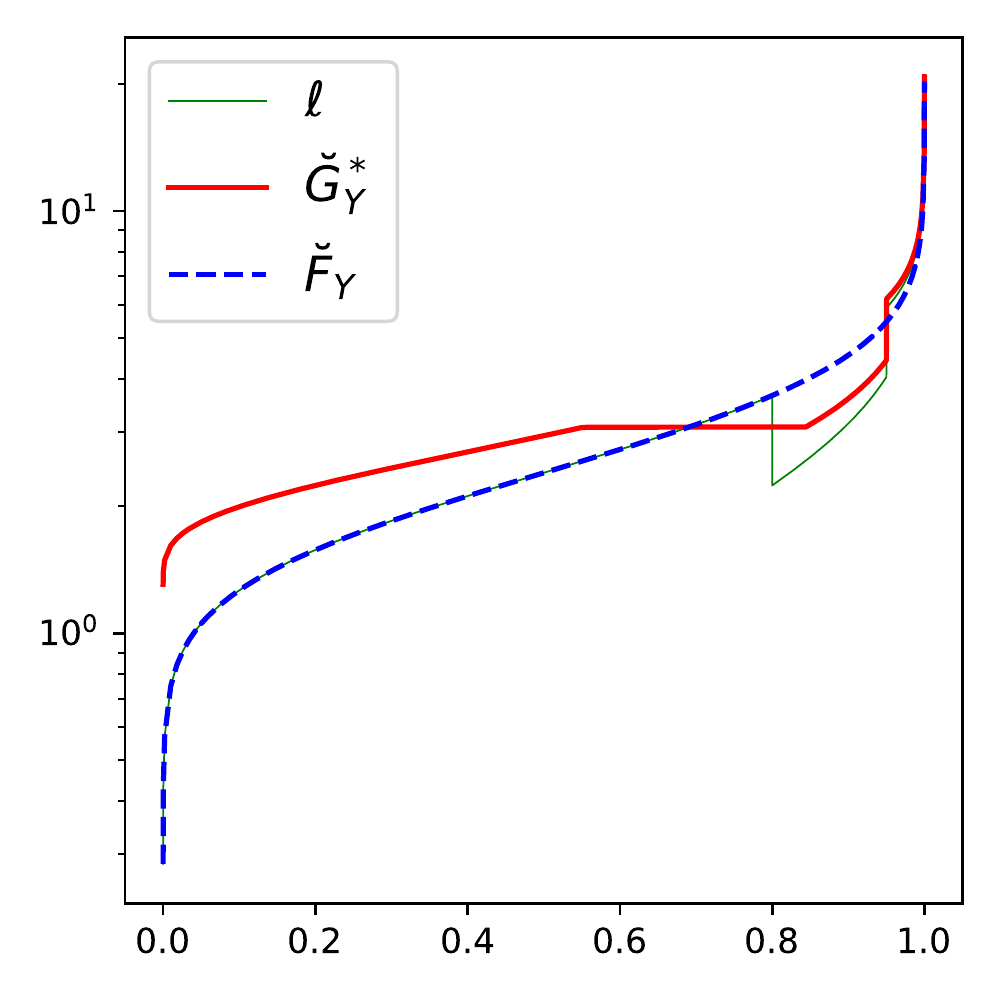}
\\[1em]
\hspace*{0.2em}
\includegraphics[width=0.31\textwidth]{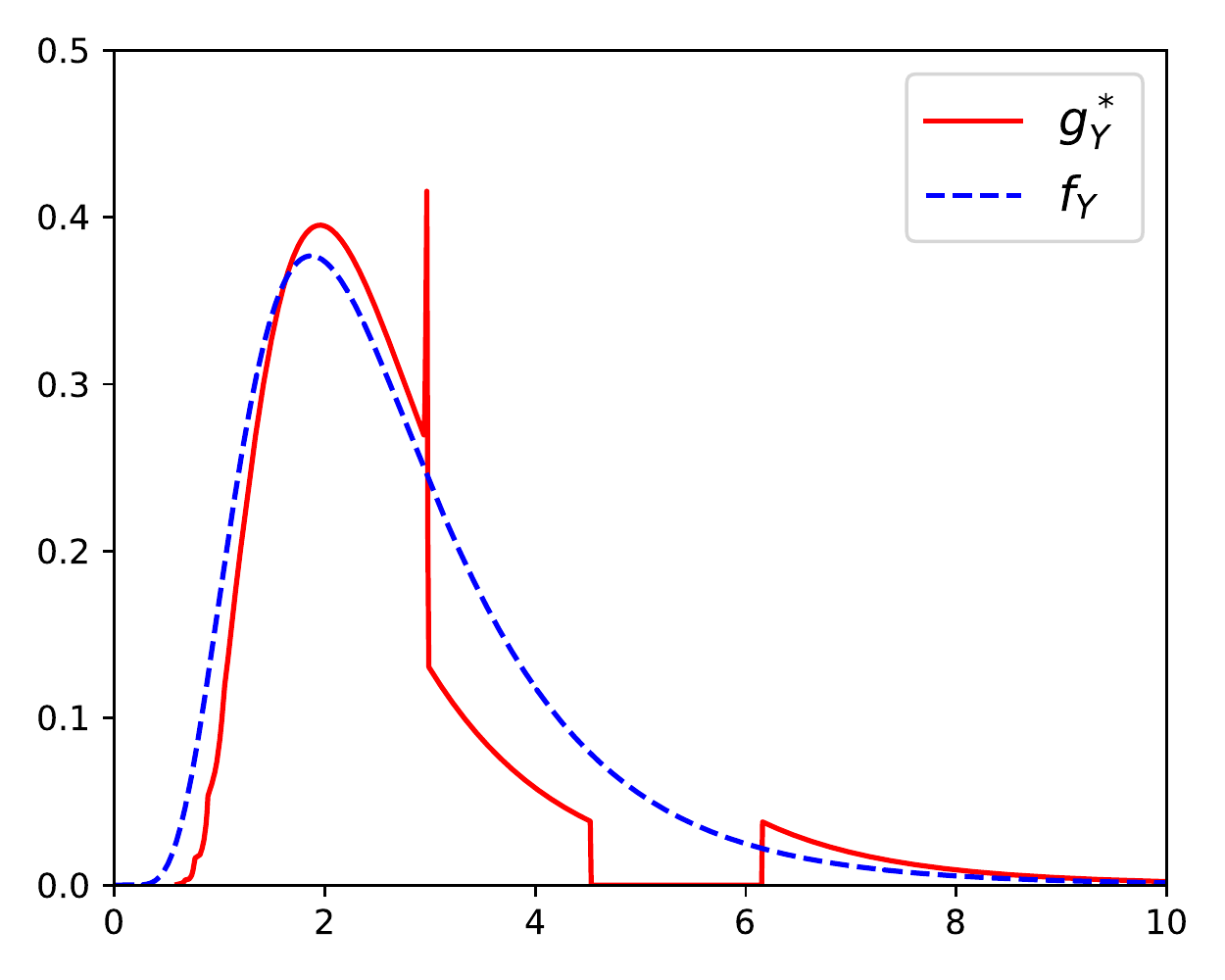}
\hspace{0.3em}
\includegraphics[width=0.31\textwidth]{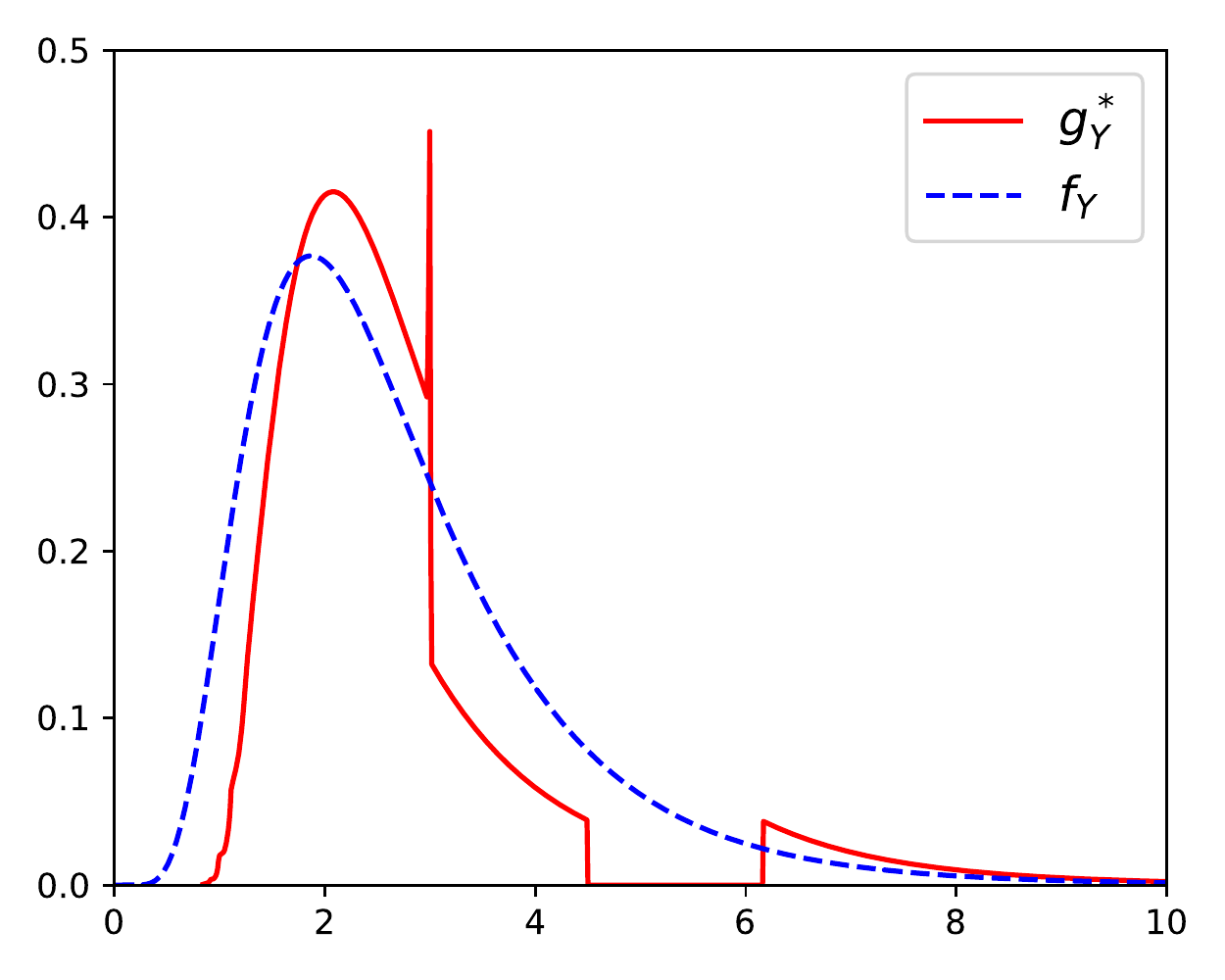}
\hspace{0.3em}
\includegraphics[width=0.31\textwidth]{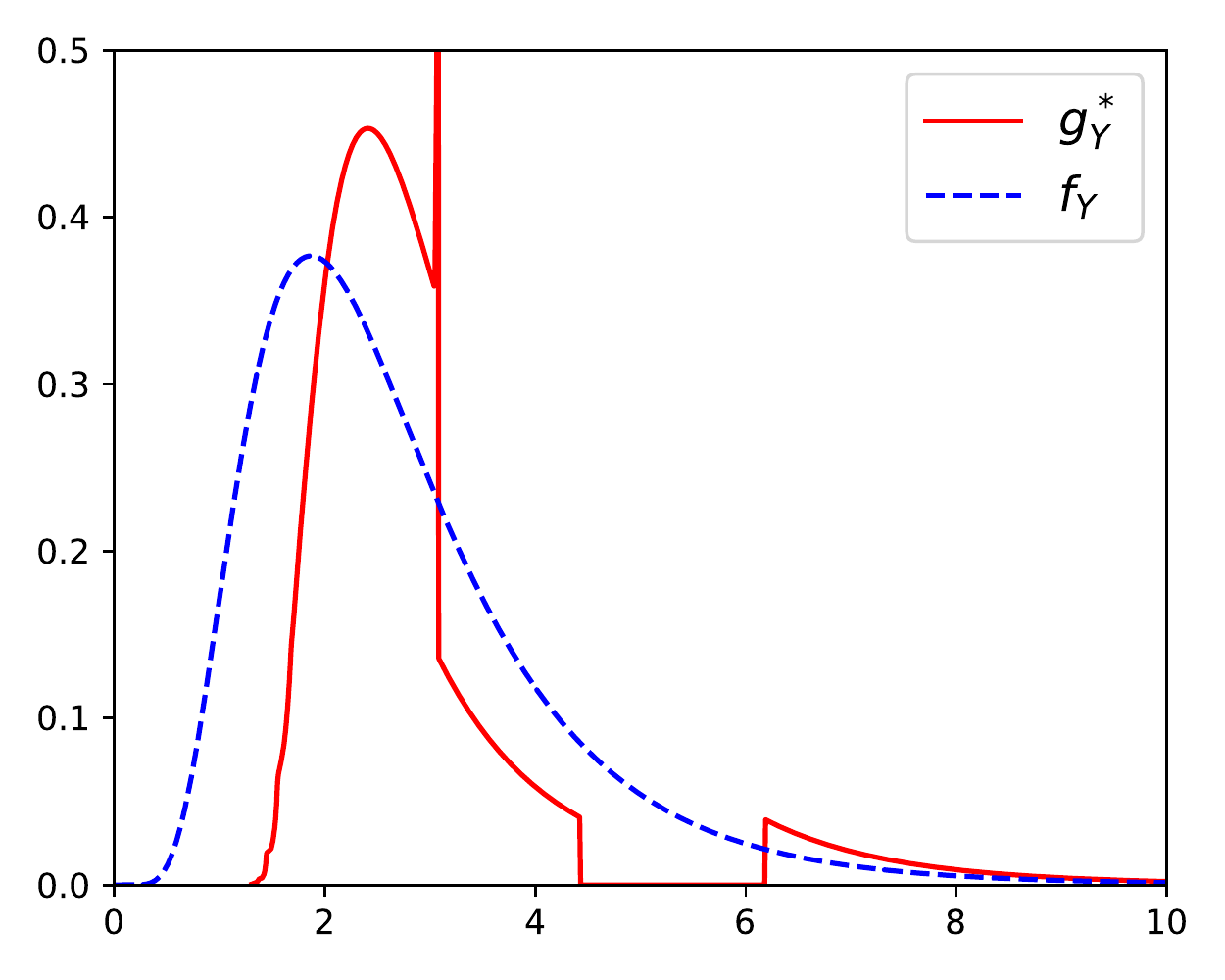}
\caption{Top panels: Baseline quantile function $\Finv_Y$ compared to the stressed quantile function $\Ginv^*_Y$, for a 10\% decrease in $\ES_{0.8}$, and a 10\% increase in $\ES_{0.95}$, and, from left to right, a 0\%, 1\%, and 3\% increase in the HARA utility, respectively. The function $\ell(\cdot)$ (solid green) is the function whose isotonic projection equals $\Ginv^*(\cdot)$. Bottom panels: Corresponding baseline $f_Y$ and stressed $g_Y^*$ densities.}
\label{fig:HARA-ES-Ginv}
\end{figure}
We again choose the baseline distribution $F_Y$ of $Y$ to be $Lognormal(\mu, \sigma^2)$ with $\mu = \frac78$ and $\sigma = 0.5$ and consider utility parameters $a = 1$, $b = 5$, and $\eta = 0.5$. Figure \ref{fig:HARA-ES-Ginv} displays the baseline and the stressed quantile functions $\Finv_Y$ and $\Ginv^*_Y$ respectively, for a combined stress on the HARA utility and on $\ES$ at levels 0.8 and 0.95. Specifically, for all three stresses we decreasing $\ES_{0.8}$ by 10\% and increasing $\ES_{0.95}$ by 10\% compared to their values under the baseline model. Moreover, the HARA utility is increased by 0\%, 1\%, and 3\%, respectively, corresponding to the panels from the left to the right. The flat part in the stressed quantile function $\Ginv^*(u)$ around $u = 0.8$, visible in all top panels of Figure \ref{fig:HARA-ES-Ginv}, is induced by the decrease in $\ES_{0.8}$ while the jump at $u = 0.95$ is due to the increase in $\ES_{0.95}$. From the left to the right panel in Figure \ref{fig:HARA-ES-Ginv}, we observe that the larger the stress on the HARA utility, the more the stressed quantile function shifts away from the baseline quantile function $\Finv_Y$.
\end{Example}

\subsection{Smoothing of the Stressed Distribution}
We observe that the stressed quantile functions derived in Section \ref{sec-opt} generally contain jumps and/or flat parts even if the baseline distribution is absolutely continuous. In situation where the this is no desirable, one may consider a smoothed version of the stressed distributions. For this we recall that the isotonic regression, the discrete counterpart of the weighted isotonic projection, of a function $\ell$ evaluated at $u_1, \ldots, u_n$ with positive weights $w_1, \ldots, w_n$, is the solution to 
\begin{equation}\label{eq:iso-regression}
    \min_{u_1, \ldots, u_n}\; \sum_{i = 1}^n \left(u_i- \ell(u_i)\right)^2 w_i\,,
    \quad \text{subject to} \quad
    u_i \le u_j\,, \quad
    i \le j\,.
\end{equation}
There are numerous efficient algorithms that solve \eqref{eq:iso-regression} most notably the pool-adjacent-violators (PAV) algorithm \cite{Barlow1972book}. It is well-known that the solution to the isotonic regression contains flat parts and jumps. A smoothed isotonic regression algorithm, termed smooth pool-adjacent-violators (SPAV) algorithm,  using an $\Lp$ regularisation was recently proposed by \cite{Sysoev2019SPAV}. Specifically, they consider 
\begin{equation*}
    \min_{u_1, \ldots, u_n}\; \sum_{i = 1}^n \left(u_i- \ell(u_i)\right)^2 w_i
    + \sum_{i = 1}^n \zeta_i \left(u_{i+1} - u_i\right)^2
    \,,
    \quad \text{subject to} \quad
    u_i \le u_j\,, \quad 
    i \le j\,,
\end{equation*}
where $\zeta_i\ge 0$, $i= 0, \ldots, n-1$, are prespecified smoothing parameters. Using a probabilitistic reasoning, \cite{Sysoev2019SPAV} argue that $\zeta_i$ may be chosen proportional to a (e.g., quadratic) kernel evaluated at $u_i$ and $u_{i+1}$, that is
\begin{equation*}
    \zeta_i = \zeta\, K(u_i, u_{i+1})\,
    \quad \text{with} \quad
    K(u_i, u_{i+1}) = \frac{1}{|u_i - u_{i+1}|^2}
    \quad \text{and} \quad
    \zeta\ge 0\,.
\end{equation*}
The choice of smoothing parameter $\zeta = 0$ correspond to the original isotonic regression larger values of $\zeta$ correspond to a greater degree of smoothness of the solution. $\zeta$ can either be prespecified or estimated using cross-validation, see also Section 3 in \cite{Sysoev2019SPAV}. 

To guarantee that the smoothed quantile function still fulfils the constraint, one may replace in every step of the optimisation for finding the Lagrange parameter the PAV with the SPAV algorithm. Thus, the Lagrange parameter are indeed found such that the constraints are fulfilled. 

\begin{Remark}
There are numerous works proposing smooth versions of isotonic regressions. Approaches include kernel smoothers, e.g., \cite{Hall2001AS}, and spline techniques, e.g., \cite{Meyer2008AAS}. These algorithms, however, are computationally heavy in that their computational cost is $O(n^2)$, where $n$ is the number of data points. Furthermore, these algorithm require a careful choice of the kernel or the spline basis which is in contrast to the SPAV. We refer the reader to \cite{Sysoev2019SPAV} for a detailed discussion and references to smooth isotonic regression algorithms.
\end{Remark}

\section{Analysing the Stressed Model} \label{sec:RN}
Recall that a modeller is equipped with a baseline model, the 3-tuple $(\X, g, \P)$, consisting of a set of input factors $\X = (X_1, \ldots, X_n)$, a univariate output random variable of interest, $Y = g(\X)$, and a probability measure $\P$. For a stress on the output's baseline distribution $F_Y$, we derived in Section \ref{sec-opt} the corresponding unique stressed distribution function, denoted here by $G^*_Y$. Thus, to fully specify the stressed model we next define a stressed probability measure $\Q^*$ that is induced by $G^*_Y$.

\subsection{The Stressed Probability Measures}
A stressed distribution $G^*_Y$ induces a canonical change of measure that allows the modeller to understand how the baseline model including the distributions of the inputs changes under the stress. The Radon-Nikodym (RN) derivative of the baseline to the stressed model is
\begin{equation*}
    \frac{d\Q^*}{d\P} = \frac{g^*_{Y}(Y)}{f_{Y}(Y)}\,,
\end{equation*}
where $f_{Y}$ and $g^*_{Y}$ denote the densities of the baseline and stressed output distribution, respectively. The RN derivative is well-defined since $f_{Y}(Y)>0$, $\P$-a.s. The distribution functions of input factors under the stressed model  -- the stressed distributions -- are then given, e.g., for input $X_i$, $i \in \{1, \ldots, n\}$, by
\begin{equation*}
    \Q^*(X_i \le x_i )  
    = \E\left[\Id_{\{X_i \le x_i\}}\, \frac{d\Q^*}{d\P} \right]\,
    = \E\left[\Id_{\{X_i \le x_i\}}\, \frac{g^*_{Y}(Y)}{f_{Y}(Y)} \right]\,,\quad x_i \in \R\,,
\end{equation*}
and for multivariate inputs $\X$ by
\begin{equation*}
    \Q^*(\X \le \x )  
    = \E\left[\Id_{\{\X \le \x\}}\, \frac{g^*_{Y}(Y)}{f_{Y}(Y)} \right]\,, \quad \x \in \R^n\,,
\end{equation*}
where $\E[\cdot]$ denotes the expectation under $\P$. Note that under the stressed probability measure $\Q^*$, the input factors' marginal and joint distributions may be altered.

\begin{Example}[HARA Utility \& ES continued]
We continue Example \ref{ex: hara-ES} and illustrate the RN-densities $\frac{d\Q^*}{d\P}$ for the following three stresses (from the left to the right panel): a 10\% decrease in $\ES_{0.8}$ and a 10\% increasing $\ES_{0.95}$ for all three stresses, and a 0\%, 1\%, and 3\% increase in the HARA utility, respectively.
\begin{figure}[b]
    \centering
\includegraphics[width=0.32\textwidth]{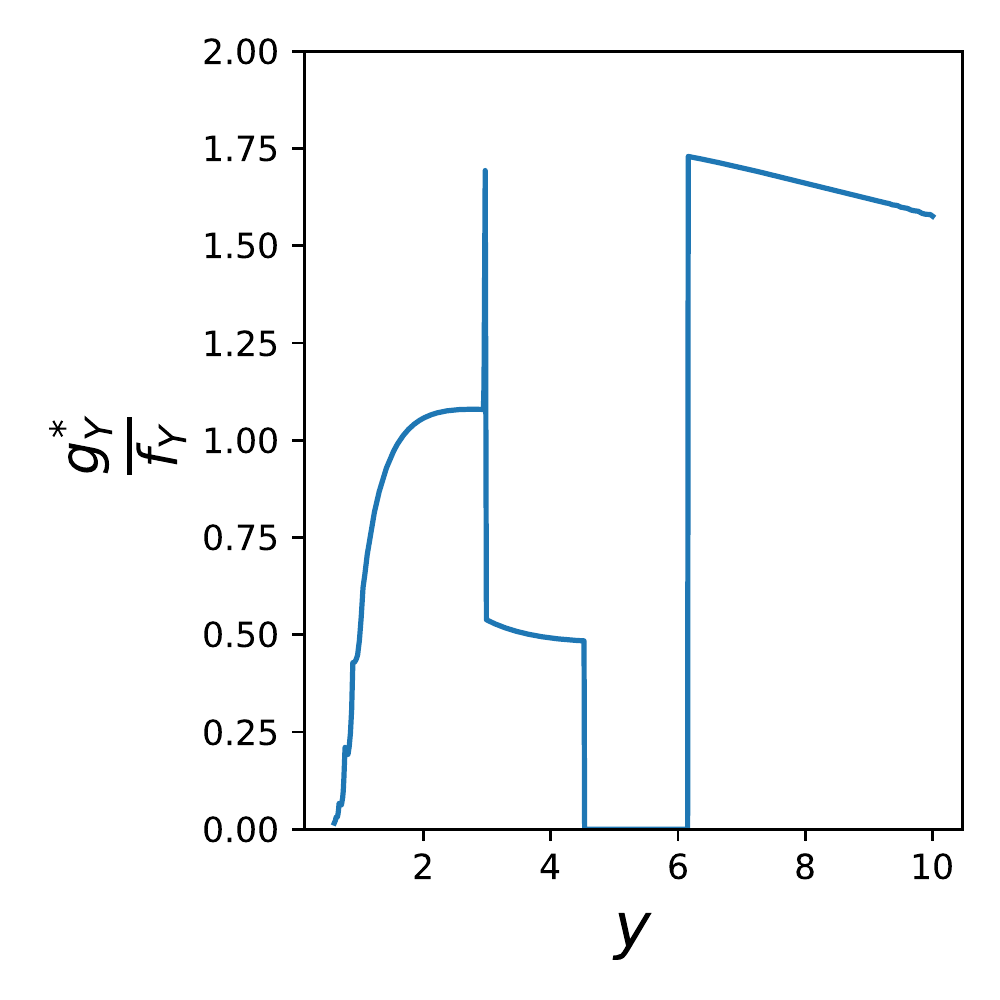}
\includegraphics[width=0.32\textwidth]{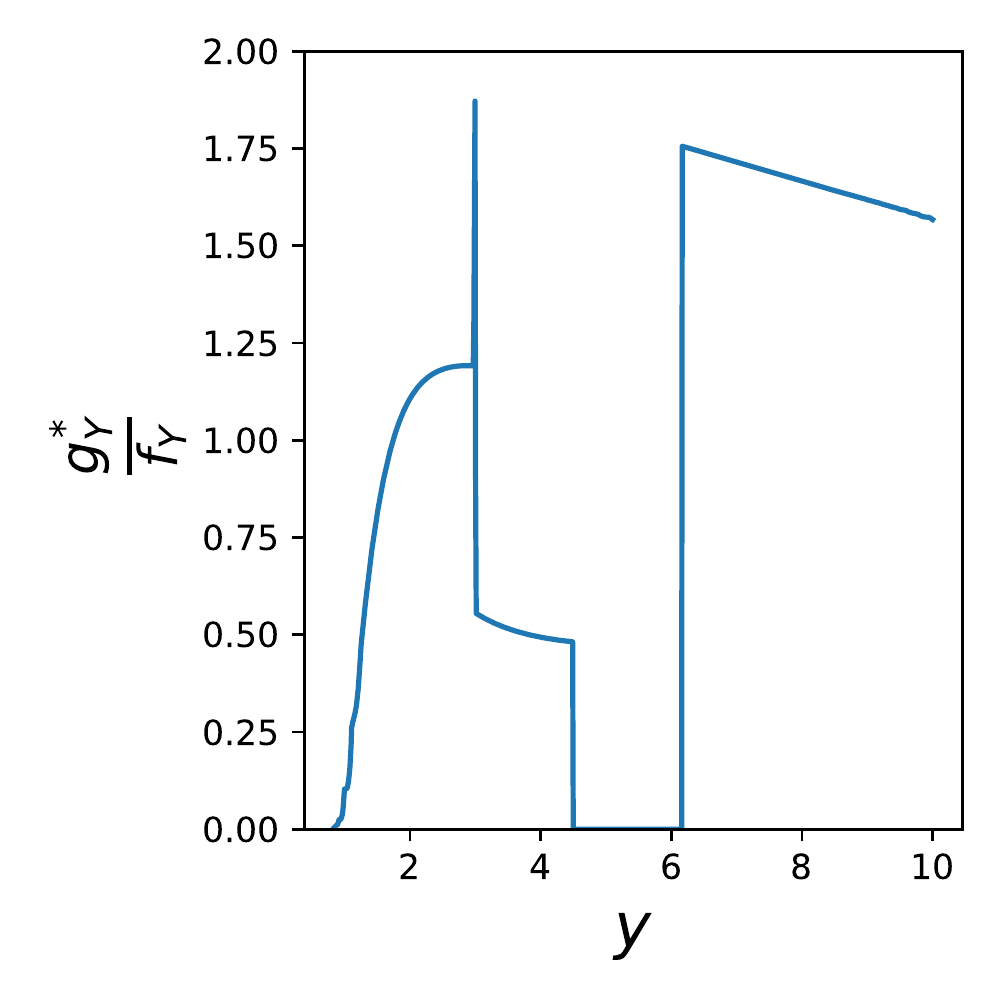}
\includegraphics[width=0.32\textwidth]{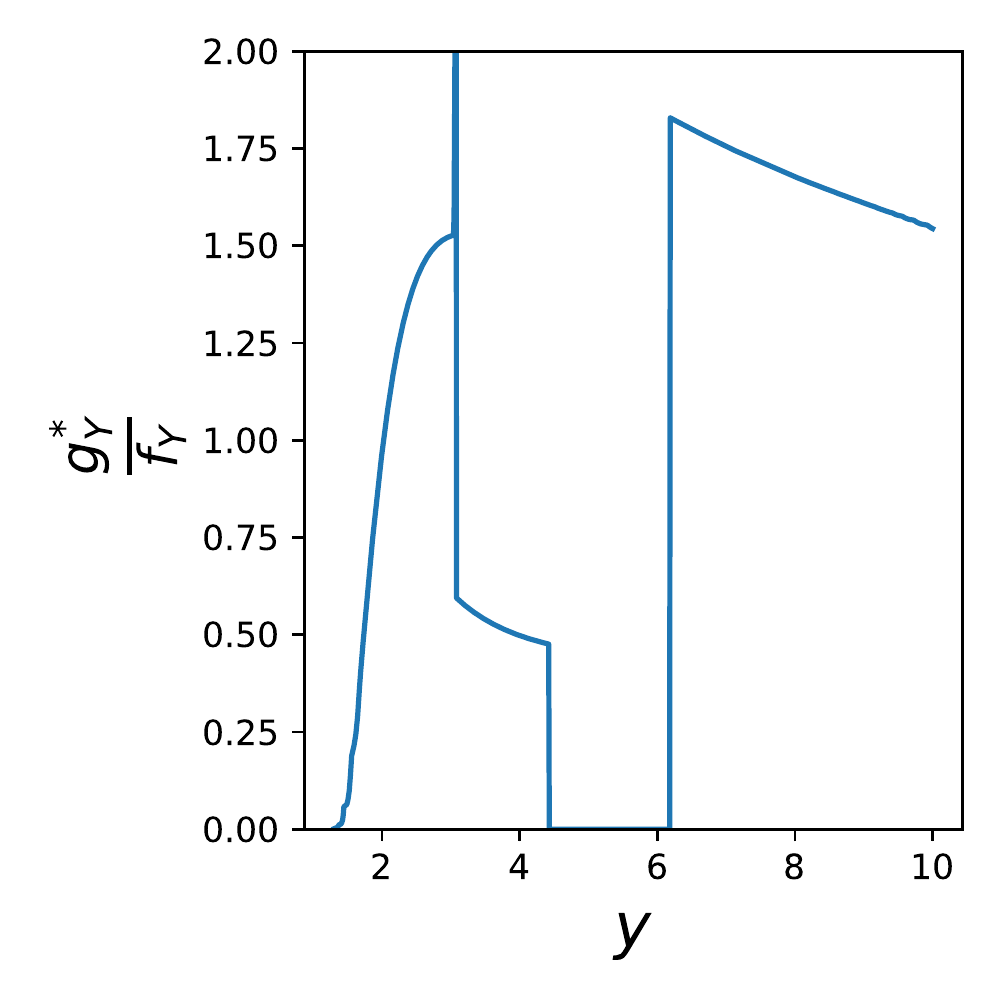}
\caption{RN-densities for following stresses: a 10\% decrease in both $\ES_{0.8}$ and $\ES_{0.95}$, and an increase in the HARA utility. The change in HARA utility is 0\%, 1\%, and 3\%, respectively, from left to right.
}
\label{fig:HARA-ES-RN}
\end{figure}%
We observe that for all three stresses large realisations of $Y$ obtain a larger weight under the stressed probability measures $\Q^*$ compared to the baseline probability $\P$. Indeed, for all three stresses it holds that $\frac{d\Q^*}{d\P}(\omega)>1$ whenever $Y(\omega) >6$ and $\omega \in \Omega$. This is in contrast to small realisations of $Y$ which obtain a weight smaller than 1. The impact of the different levels of stresses of the HARA utility (0\%, 1\%, and 3\%, from the left to the right panel) can be observed in the left tail of $\frac{d\Q^*}{d\P}$; a larger stress on the utility induces larger weights. The length of the trough of $\frac{d\Q^*}{d\P}$ is increasing from the left panel (approx. ($4.53, 6.15$)) to the right panel (approx. ($4.43, 6.18$)), and corresponds in all cases to the constant part in $G^*_Y$ (see Figure \ref{fig:HARA-ES-Ginv}, top panels) which is induced by the decrease in $\ES_{0.8}$ under the stressed model. 
\end{Example}

\subsection{Reverse Sensitivity Measures}
Comparison of the baseline and a stressed model can be conducted via different approaches depending on the modeller's interest. While probabilistic sensitivity measures underlie the assumption of a fixed probability measure and quantify the divergence between the conditional (on a model input) and the unconditional output density \cite{Saltelli2008book}, the proposed framework compares a baseline and a stressed model, i.e., distributions under different probability measures. Therefore, to quantify the distributional change in input factor $X_i$ from the baseline $\P$ to the stressed $\Q^*$ probability, a sensitivity measure introduced by \cite{Pesenti2019EJOR} may be suitable which quantifies the variability of an input factor's distribution from the baseline to the stressed model. A generalisation of the reverse sensitivity measure is stated here.

\begin{Definition}[Marginal Reverse Sensitivity Measure \cite{Pesenti2019EJOR}]\label{def:reverse-sens}
For a function $s \colon \R \to \R$, the reverse sensitivity measure to input $X_i$ with respect to a stressed probability measure $\Q^*$ is defined by
\begin{align*}
    \S_i^{\Q^*} = 
    \begin{cases}
    \quad \cfrac{\E^{\Q^*} [s(X_i) ] - \E[s(X_i)]}{ \max\limits_{\Q \in \mQ}\E^{\Q} [s(X_i)] - \E[s(X_i)]}
        \qquad\qquad &  \E^{\Q^*} [s(X_i)] \ge \E[s(X_i)]\,,
    \\[1em]
    -\;\cfrac{\E^{\Q^*} [s(X_i) ] - \E[s(X_i)]}{ \min\limits_{\Q \in \mQ}\E^{\Q} [s(X_i)] - \E[s(X_i)]}
        & \text{otherwise,}
    \end{cases}
\end{align*}
where $\mQ = \{ \Q \;| \; \Q \quad \text{probability measure with} \quad \frac{d\Q}{d\P}\stackrel{\P}{=} \frac{d\Q^*}{d\P}\}$ is the set of all probability measures whose RN-derivative have the same distribution as $\frac{d\Q^*}{d\P}$ under $\P$. We adopted the convention that $\pm\frac{ \infty}{\infty} = \pm 1$ and $\frac{0}{0} = 0$. 
\end{Definition}
The sensitivity measure is called ``reverse'', as the stress is applied to the output random variable $Y$ and the sensitivity monitors the change in input $X_i$. Definition \ref{def:reverse-sens} applies, however, also to stresses on input factors, in which case the RN-density $\frac{d\Q^*}{d\P}$ is a function of the stressed input factor and we refer to \cite{Pesenti2019EJOR} for a discussion. Note, that the reverse sensitivity measure can be viewed as a normalised covariance measure between the input $s(X_i)$ and the Radon Nikodym derivative $\frac{d\Q^*}{d\P}$. 

The next proposition provides a collection of properties that the reverse sensitivity measure possesses, we also refer to \cite{Pesenti2019EJOR} for a detailed discussion of these properties. For this we first recall the definition of comonotonic and counter-monotonic random variables.

\begin{Definition}
Two random variables $Y_1$ and $Y_2$ are comonotonic under $\P$, if and only if, there exists a random variable $W$ and non-decreasing functions $h_1, h_2\colon \R \to \R$, such that the following equalities hold in distribution under $\P$
\begin{equation}\label{eq:como}
    Y_1 = h_1(W) \quad \text{and} \quad Y_2 = h_2(W).
\end{equation}
The random variables $Y_1$ and $Y_2$ are counter-monotonic under $\P$, if and only if, \eqref{eq:como} holds with one of the functions $h_1(\cdot), h_2(\cdot)$ being non-increasing, and the other non-decreasing. 
\end{Definition}

If two random variables are (counter) comonotonic under one probability measure, then they are also (counter) comonotonic under any other absolutely continuous probability measure, see e.g., Proposition 2.1 of \cite{Cuestaalbertos1993JMA}. Thus, we omit the specification of the probability measure when discussing counter- and comonotonicity.

\begin{Proposition}[Properties of Reverse Sensitivity Measure]\label{prop:properties-rev-sens}
The reverse sensitivity measure possesses the following properties:
\begin{enumerate}[label = \roman*)]

    \item $\S_i^{\Q^*}\in [-1,1]$;
    
    \item $\S_i^{\Q^*} = 0$ if $(s(X_i), \frac{d\Q^*}{d\P})$ are independent under $\P$;
    
    \item $\S_i^{\Q^*} = 1$ if and only if $(s(X_i), \frac{d\Q^*}{d\P})$ are comonotonic; 
    
    \item $\S_i^{\Q^*} = -1$ if and only if $(s(X_i), \frac{d\Q^*}{d\P})$ are counter-comonotonic.
\end{enumerate}
\end{Proposition}
The function $s(\cdot)$ provides the flexibility to create sensitivity measures that quantify changes in moments, e.g., via $s(x) = x^k$, $k \in \N$, or in the tail of distributions, e.g., via $s(x) = \Id_{\{x > \VaR_\alpha(X_i)\}}$, for $\alpha \in (0,1)$.

Next, we generalise Definition \ref{def:reverse-sens} to a sensitivity measure that accounts for multiple input factors. While $\S_i^{\Q^*}$ measures the change of the distribution of $X_i$ from the baseline to the stressed model the sensitivity $\S_{i,j}^{\Q^*}$ introduced below, quantifies how the joint distribution of $(X_i, X_j)$ changes when moving from $\P$ to $\Q^*$.

\begin{Definition}[Bivariate Reverse Sensitivity Measure]
For a function $s \colon \R^2 \to \R$, the reverse sensitivity measure to inputs $(X_i,X_j)$ with respect to a stressed probability measure $\Q^*$ is defined by
\begin{align*}
    \S_{i,j} ^{\Q^*} = 
    \begin{cases}
    \quad\cfrac{\E^{\Q^*} [s(X_i, X_j) ] - \E[s(X_i, X_j)]}{ \max\limits_{\Q \in \mQ}\E^{\Q} [s(X_i, X_j)] - \E[s(X_i, X_j)]}
        \qquad\qquad &  \E^{\Q^*} [s(X_i, X_j)] \ge \E[s(X_i, X_j)]\,,
    \\[1em]
    -\;\cfrac{\E^{\Q^*} [s(X_i, X_j) ] - \E[s(X_i, X_j)]}{ \min\limits_{\Q \in \mQ}\E^{\Q} [s(X_i, X_j)] - \E[s(X_i, X_j)]}
        & \text{otherwise}\,,
    \end{cases}
\end{align*}
where $\mQ$ is given in Definition \ref{def:reverse-sens}.
\end{Definition}
The bivariate sensitivity measure satisfies all the properties in Proposition \ref{prop:properties-rev-sens} when $s(X_i)$ is replaced by $s(X_i, X_j)$. The bivariate sensitivity $\S_{i,j}^{\Q^*}$ can also be generalised to $k$ input factors by choosing a function $s\colon \R^k \to \R$.

\begin{Remark}
Probabilistic sensitivity measures are typically used for importance measurement and take values in $[0,1]$; with 1 being the most important input factor and 0 being (desirably) independent from the output \cite{Borgonovo2021EJOR}. This is in contrast to our framework where $\S_i^{\Q^*}$ lives in $[-1,1]$ and e.g., a negative dependence, such as negative quadrant dependence between $s(X_i)$ and $\frac{d\Q^*}{d\P}$ implies that $\S_i^{\Q^*} <0$, see  \cite{Pesenti2019EJOR}[Proposition 4.3]. Thus, the proposed sensitivity measure is different in that it allows for negative sensitivities where the sign of $\S_i^{\Q^*}$ indicates the direction of the distributional change. 
\end{Remark}

\section{Application to a Spatial Model} \label{sec:application}
We consider a spatial model for modelling insurance portfolio losses where each individual loss occurs at different locations and the dependence between individual losses is a function of the distance between the locations of the losses. Mathematically, denote the locations of the insurance losses by $\z_1, \ldots, \z_{10}$, where $\z_m = (z_m^1, z_m^2) $ are the coordinates of location $\z_m$, $m = 1, \ldots, 10$. The insurance loss at location $m$, denoted by $L_m$, follows a $Gamma(5, \frac{0.2}{m})$ distribution with location parameter $25$. Thus, the minimum loss at each location is 25 and locations with larger mean also exhibit larger standard deviations. The losses $L_1, \ldots, L_m$ have, conditionally on $\Theta = \theta$, a Gaussian copula with correlation matrix given by $\rho_{i,j} = \Cor(L_i, L_j) = e^{-\theta ||\z_i - \z_j||}$, where $|| \cdot ||$ denotes the Euclidean distance. Thus, the further apart the locations $\z_i$ and $\z_j$ are the smaller the correlation between $L_i$ and $L_j$. The parameter $\Theta$ takes values $(0, 0.4, 5)$ with probabilities $(0.05, 0.6, 0.35)$ that represent different regimes. Indeed $\Theta = 0$ corresponds to a correlation of 1 between all losses, independently of their location. Larger values of $\Theta$ correspond to smaller albeit still positive correlation. Thus, regime with $\Theta = 0$ can be viewed as, e.g., circumstances suitable for natural disasters. We further define the total loss of the insurance company by $Y = \sum_{m = 1}^{10} L_m$. 

We perform two different stresses on the total loss $Y$ detailed in Table \ref{tab:stresses}. Specifically, we consider as a first stress a 0\% change in HARA utility, a 0\% change in $\ES_{0.8}(Y)$, and a 1\% increase in $\ES_{0.95}(Y)$ from the baseline to the stressed model. The second stress is composed of a 1\% increase in HARA utility, a 1\% increase in $\ES_{0.8}(Y)$, and a 3\% increase in $\ES_{0.95}(Y)$ compared to the baseline model. As the second stress increases all three metrics it may be viewed as a more severe distortion of the baseline model. 
\begin{table}[htbp]
  \centering
  \caption{Summary of the stresses applied to the portfolio loss $Y$ represented in relative increases of the stressed model from the baseline model.}
\begin{tabular}{lccc}
\toprule
      &   HARA utility &   $\ES_{0.8}(Y)$ &  $\ES_{0.95}(Y)$\\
    \textbf{Stress 1:} $\Q^*_1$ &  0\% &  0\% & 1\%\\
    \textbf{Stress 2:} $\Q^*_2$ & 1\%  &  1\% & 3\%\\
    \bottomrule
    \end{tabular}%
  \label{tab:stresses}%
\end{table}

\begin{figure}[h]
    \centering
\includegraphics[width=0.32\textwidth]{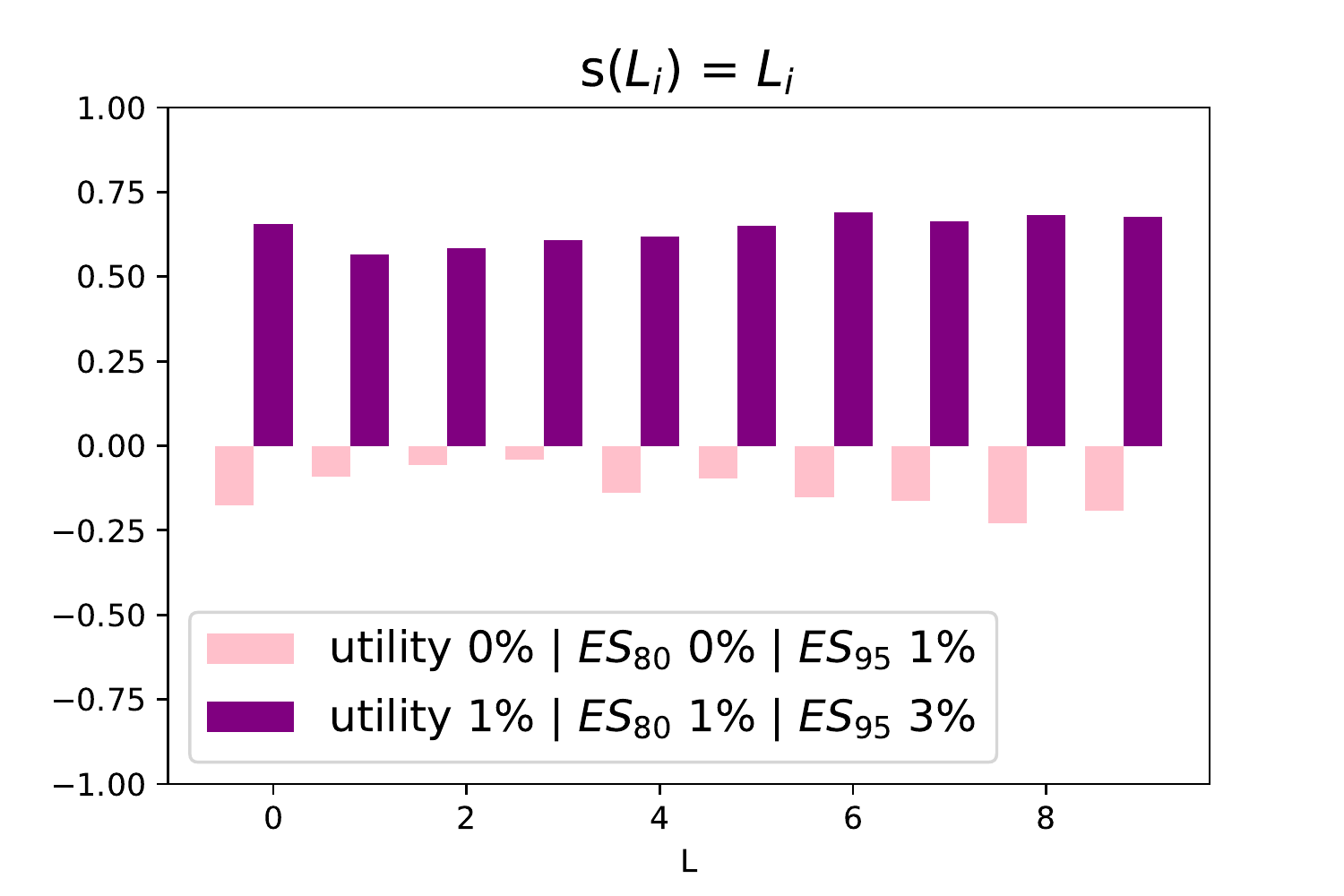}
\includegraphics[width=0.32\textwidth]{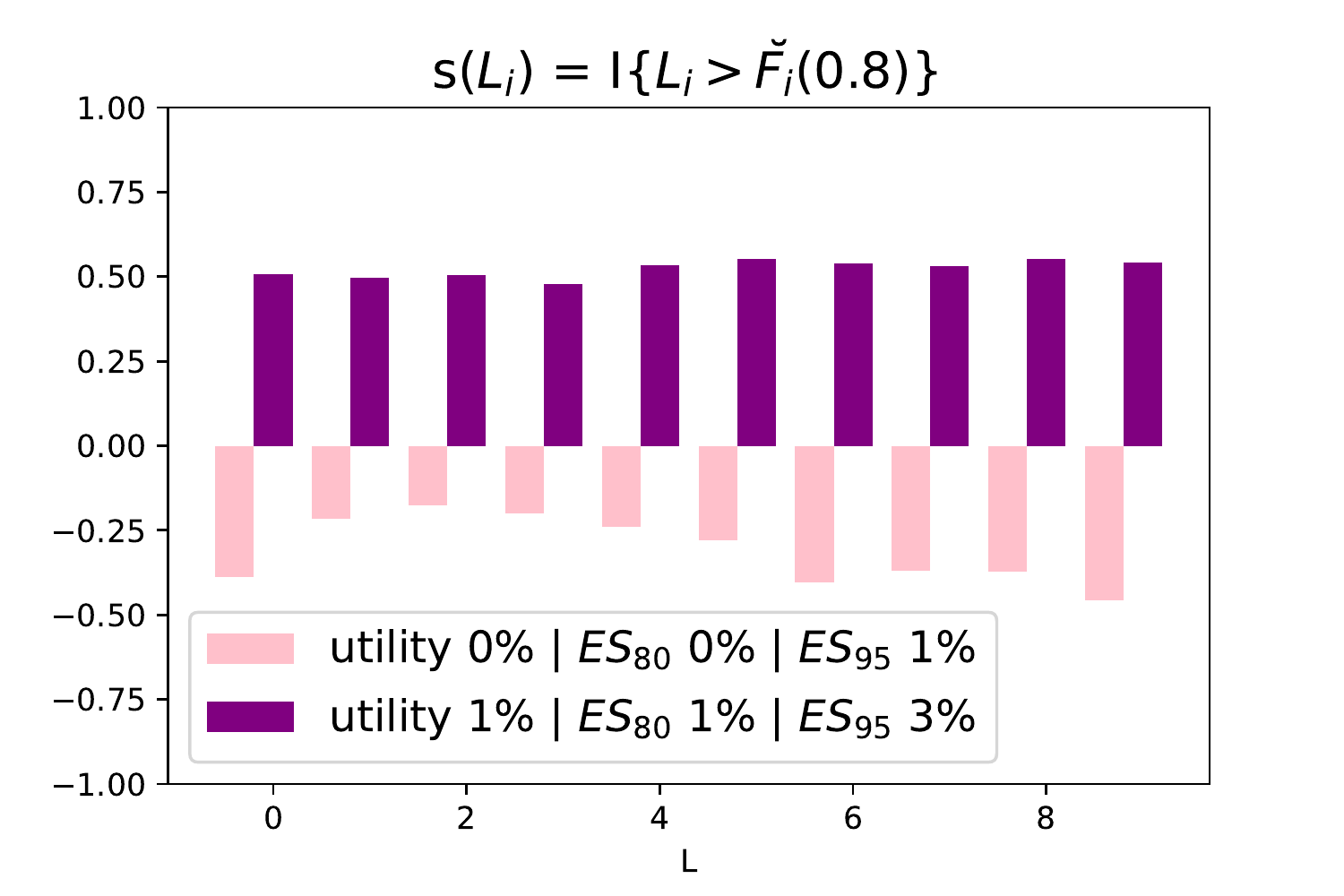}
\includegraphics[width=0.32\textwidth]{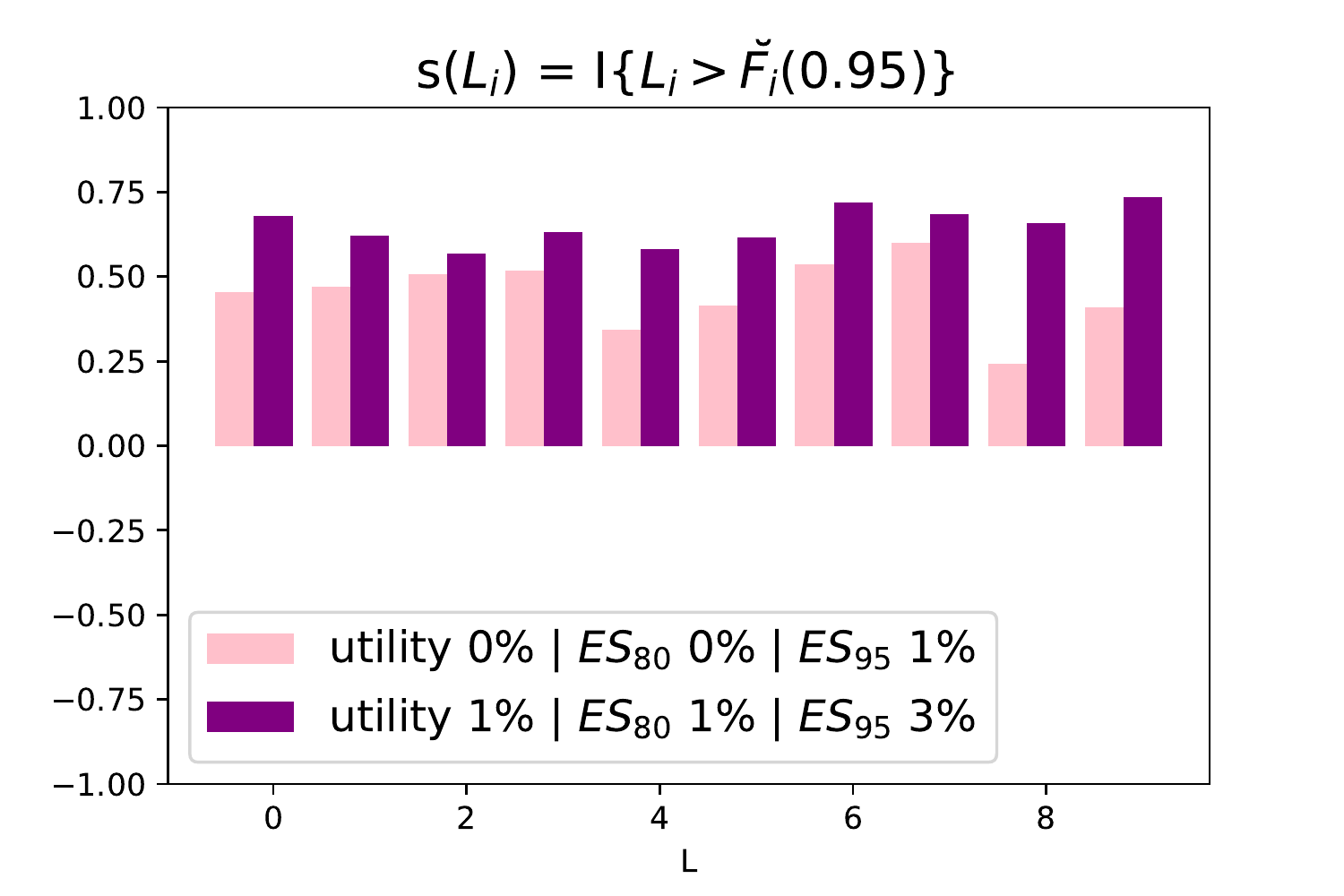}
\caption{Reverse sensitivity measures with $s(x) = x$, $s(x) = \Id_{\{x > \Finv_i(0.8)\}}$, and $s(x) =  \Id_{\{x > \Finv_i(0.95)\}}$ (left to right), for two different stresses on the output $Y$. First stress (salmon) is keeping the HARA utility and $\ES(Y)_{0.8}$ fixed and increasing the $\ES(Y)_{0.85}$ by 1\%. Second stress (violet) is an increase of 1\% in HARA utility, 1\% $\ES(Y)_{0.8}$, and 3\% in $\ES(Y)_{0.85}$.}
\label{fig:HARA-ES-RN-reverse-sens}
\end{figure}

Next, we calculate reverse sensitivity measures for the losses $L_1 , \ldots, L_{10}$ for both stresses $\Q^*_1$ and $\Q^*_2$. Figure \ref{fig:HARA-ES-RN-reverse-sens} displays the reverse sensitivity measures for functions $s(x) = x$, $s(x) = \Id_{\{x > \Finv_i(0.8)\}}$, and $s(x) =  \Id_{\{x > \Finv_i(0.95)\}}$, from the left to the right panel, and where $\Finv_i$, denotes the $\P$-quantile function of $L_i$, $i = 1, \ldots, 10$. 
We observe that for stress 2, the reverse sensitivities to all losses $L_i$ and all choices of function $s(\cdot)$ are positive. This contrasts the reverse sensitivities for stress 1. Indeed, for stress 1 the reverse sensitivities with both $s(x) = x$ and $s(x) = \Id_{\{x > \Finv_i(0.8)\}}$ are negative, with the former values being smaller indicating a smaller change in the distributions of the $L_i$'s. By definition of the reverse sensitivity, the left panel corresponds to the (normalised) difference between the expectation under the stressed and baseline model. The middle and right panels correspond to the (normalised) change in the probability of exceeding $\Finv_i(0.8)$ and $\Finv_i(0.95)$, respectively. Thus, as seen in the plots, while the expectations and probabilities of exceeding the 80\% $\P$-quantile are smaller under the stressed model, the probabilities of exceeding the 95\% $\P$-quantile are increased substantially. The first stress increases the $\ES$ at level 0.95 while simultaneously fixes the utility and $\ES$ at level 0.8 to its values under the baseline model. This induces under the stressed probability measure a reduction of the mean and of the probability of exceeding the 80\% $\P$-quantile while the probability of exceeding the 95\% $\P$-quantile increases. Thus, the reverse sensitivity measures provide a spectrum of measures to analyse the distributional change of the losses $L_i$ from the baseline to the stressed model. 

Next, for a comparison we calculate the delta sensitivity measure of introduced by \cite{Borgonovo2007RESS}. For a probability measure $\Q$  the delta measure of $L_i$ is defined by
\begin{equation*}
    \xi^{\Q}(L_i) = \tfrac12 \int\,\int\, \left| f_Y^\Q(y) - f^\Q_{Y|i}(y\, |\, z)\right|\, f_i^\Q(z) \, dy\, dz,
\end{equation*}
where $f_Y^\Q(\cdot) $ and $f^\Q_{i}(\cdot)$ are the densities of $Y$ and $L_i$ under $\Q$, respectively, and where $f^\Q_{Y|i}(\cdot|\cdot)$ is the conditional density of the total portfolio loss $Y$ given $L_i$ under $\Q$. 
\begin{table}[t]
  \centering
  \caption{Comparison of different sensitivity measures: First two columns correspond to the reverse sensitivity measures with $s(x) = \Id_{\{x > \Finv(0.95)\}}$ and stressed models $\Q_1^*$, and  $\Q_2^*$, respectively. The last three columns are the delta measure under $\P$, $\Q_1^*$, and  $\Q_2^*$, respectively}
\begin{tabular}{lccccccc}
\toprule
      &   &  $\S_i^{\Q^*_1}$ & $\S_i^{\Q^*_2}$ &   & $\xi^{\P}$ & $\xi^{ \Q_1^*}$ &  $\xi^{ \Q_2^*}$ \\[0.5em]
    $L_1$ &   &  0.45 & 0.68 &   & 0.38 & 0.38 &   0.38 \\
    $L_2$ &   &  0.47 & 0.62 &   & 0.29 & 0.29 &   0.29 \\
    $L_3$ &   &  0.51 & 0.57 &   & 0.30 & 0.30 &   0.29 \\
    $L_4$ &   &  0.52 & 0.63 &   & 0.30 & 0.30 &   0.29 \\
    $L_5$ &   &  0.34 & 0.58 &   & 0.33 & 0.34 &   0.33 \\
    $L_6$ &   &  0.41 & 0.62 &   & 0.34 & 0.34 &   0.32 \\
    $L_7$ &   &  0.54 & 0.72 &   & 0.40 & 0.40 &   0.38 \\
    $L_8$ &   &  0.60 & 0.69 &   & 0.38 & 0.39 &   0.39 \\
    $L_9$ &   &  0.24 & 0.66 &   & 0.40 & 0.40 &   0.38 \\
 $L_{10}$ &   &  0.41 & 0.73 &   & 0.39 & 0.38 &   0.37 \\
    \bottomrule
    \end{tabular}%
  \label{tab:sensitivities}%
\end{table}%
Table \ref{tab:sensitivities} reports the delta measures under the baseline model $\xi^{\P}$ and the two stresses, i.e. $\xi^{\Q^*_1}$ and $\xi^{\Q^*_2}$. We observe that the delta measures are similar for all losses $L_i$ and do not change significantly under the different probability measures. As the delta sensitivity measure quantifies the importance of input factors under a probability measure, having similar values for $\xi^\P$, $\xi^{\Q^*_1}$, and $\xi^{\Q^*_1}$, means that the importance ranking of the $L_i$'s under different stresses does not change. We also report, in the first two columns of Table \ref{tab:sensitivities}, the reverse sensitivity measures with $s(x) = \Id_{\{x > \Finv(0.95)\}}$. The reverse sensitivity measures provide, in contrast to the delta measure, insight into the change in the distributions of the $L_i$'s from $\P$ and $\Q^*$.

\begin{figure}[h]
    \centering
\includegraphics[width=0.49\textwidth]{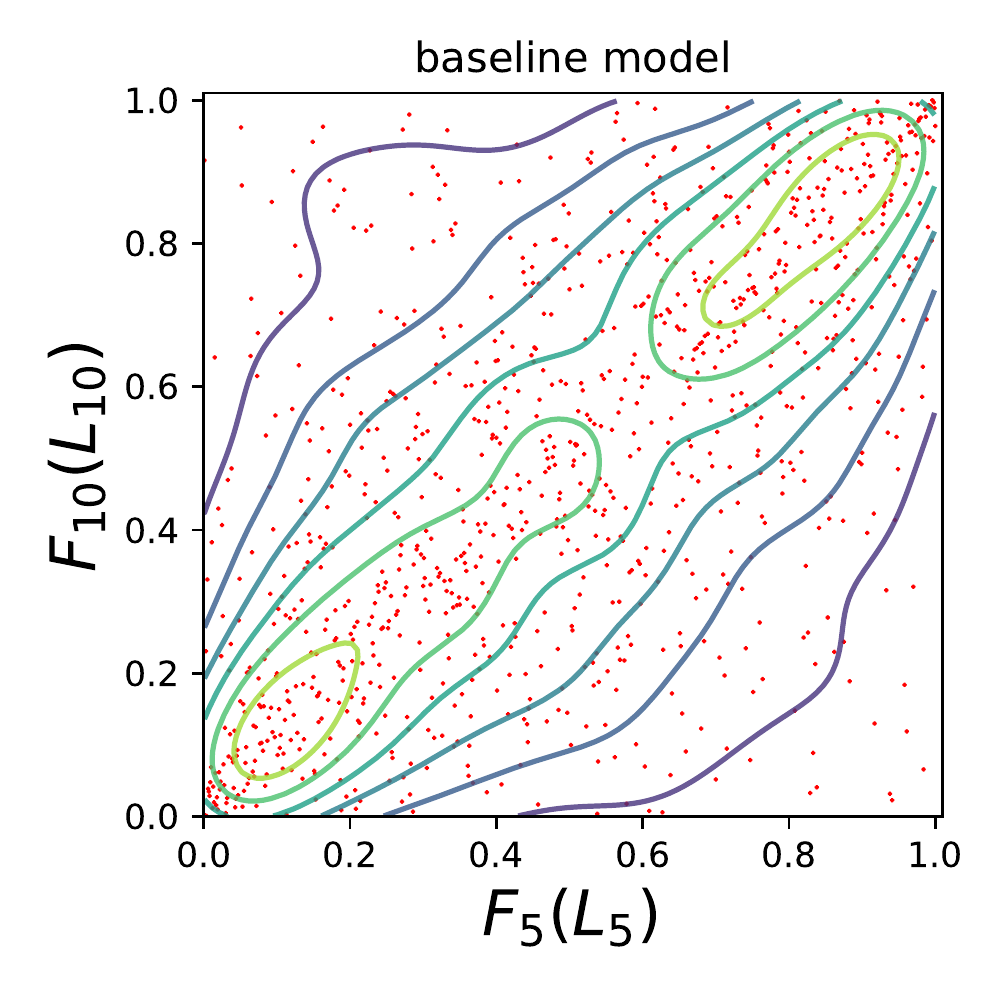}
\includegraphics[width=0.49\textwidth]{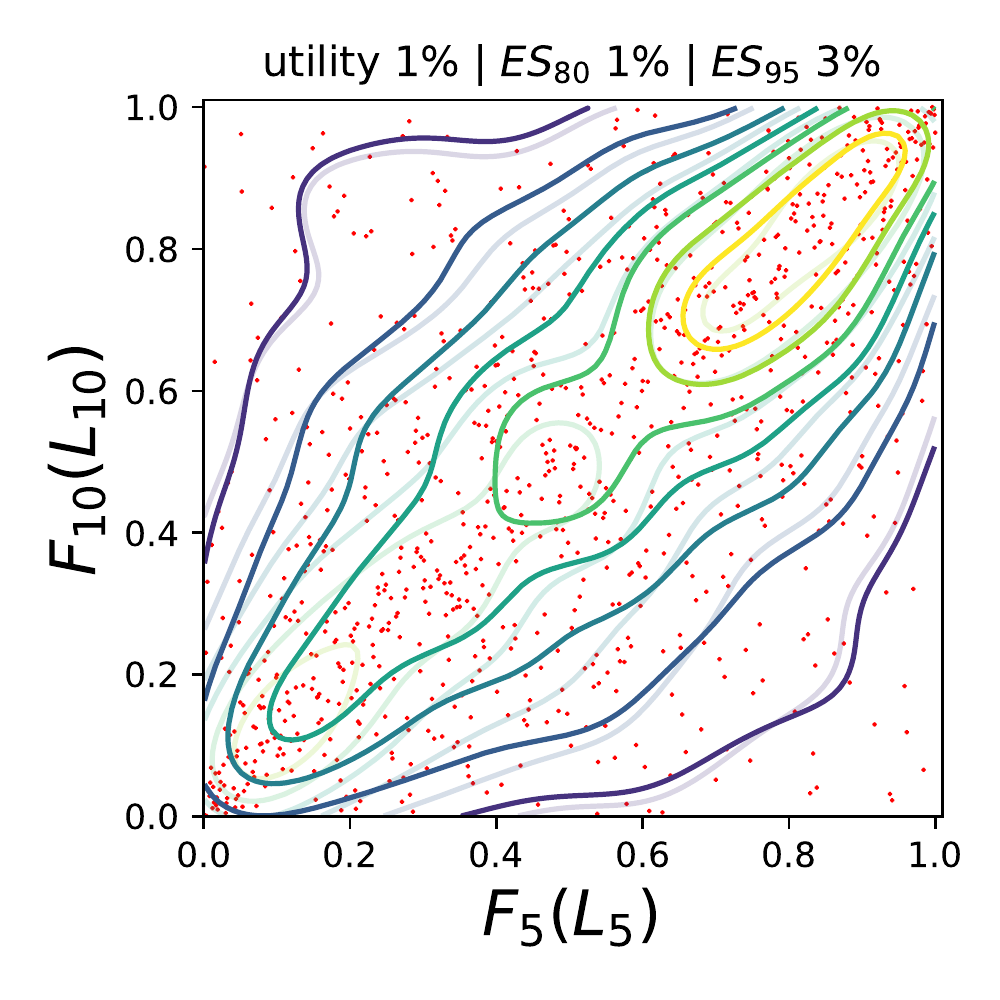}\\
\includegraphics[width=0.49\textwidth]{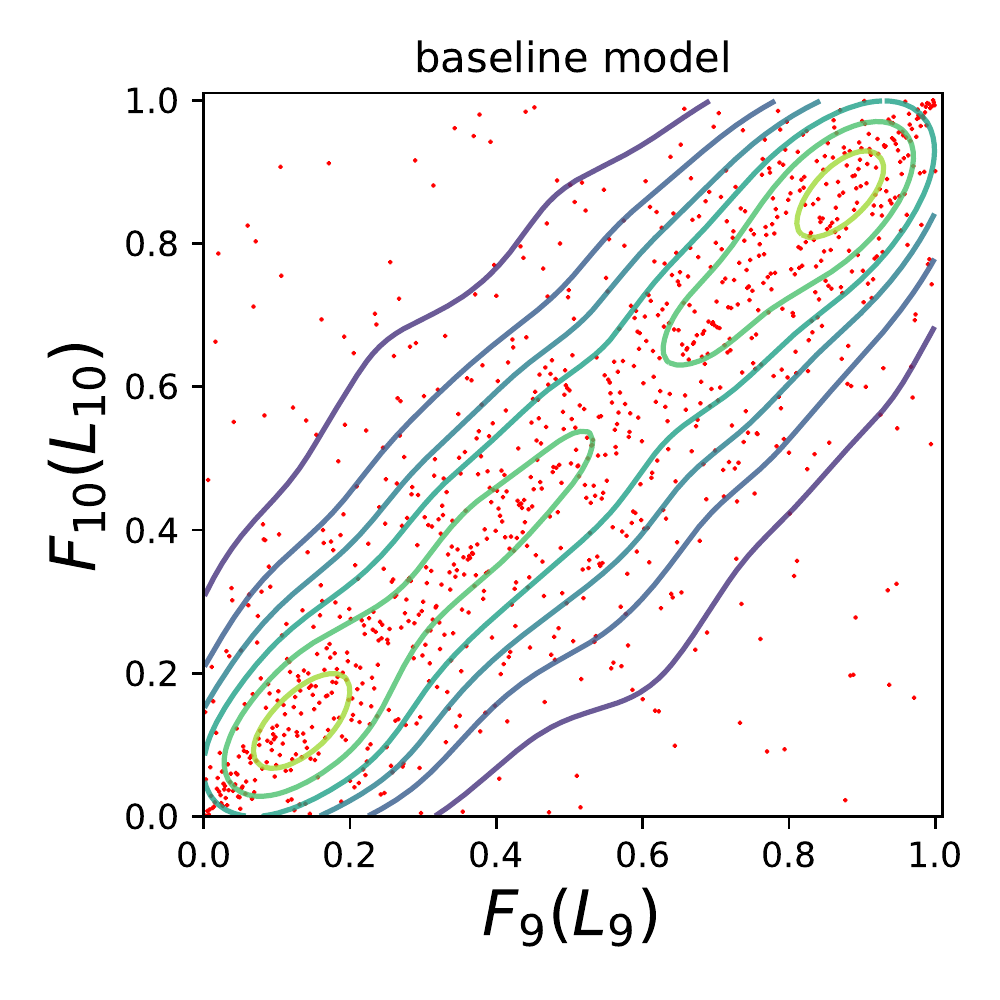}
\includegraphics[width=0.49\textwidth]{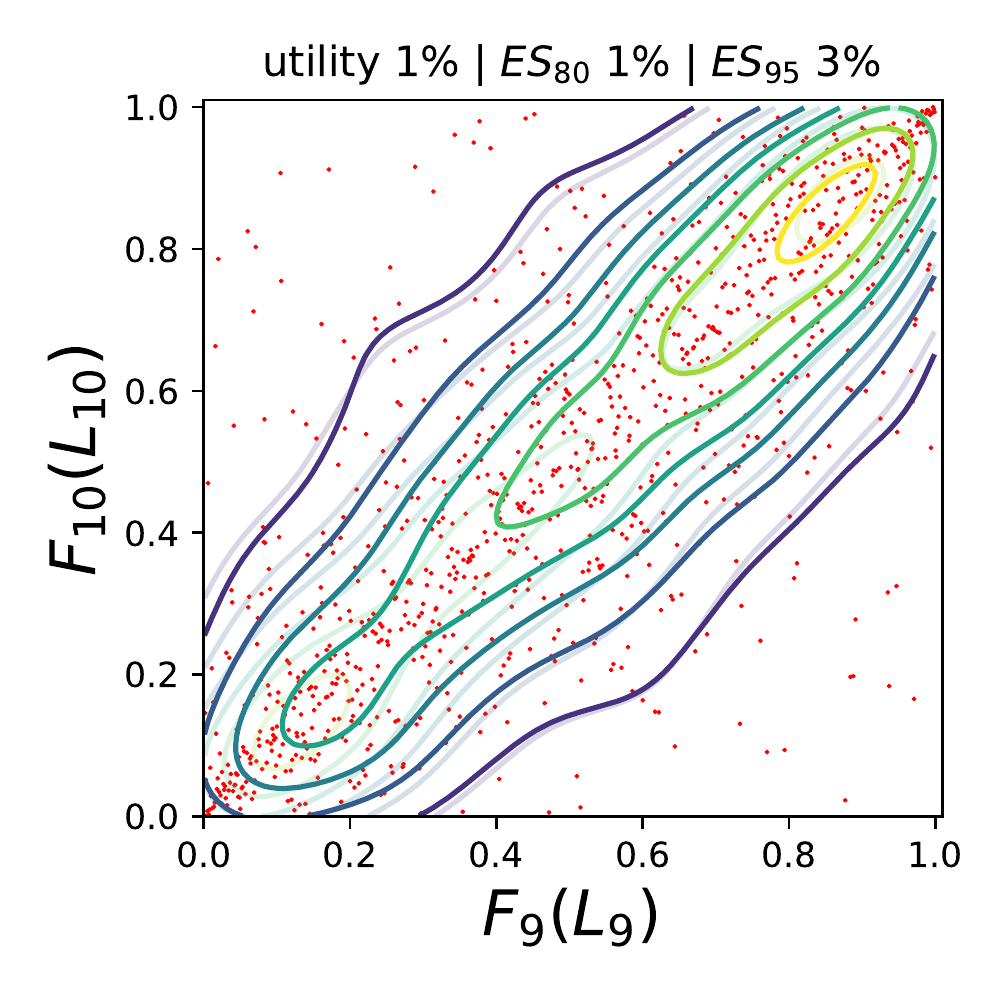}
\caption{Contour plots of the bivariate copulae of $(L_5, L_{10})$ (top panels) and $(L_9, L_{10})$ (bottom panels) under different models. The left contour plots correspond to the baseline model and the right panels to the stress $\Q^*_2$ (solid lines) with the baseline contours reported using partially transparent lines. Red points are simulated realisations.}
\label{fig:HARA-ES-contour}
\end{figure}%
Alternatively to considering the change in the marginal distributions $L_i$ from the baseline to the stressed model, we can study the change in the dependence between the losses when moving from the baseline to a stressed model. For this, we consider the  bivariate reverse sensitivity measures for the pairs $(L_5, L_{10})$ and $(L_9, L_{10})$ for the second stress $\Q^*_2$, that is a 1\% increase in HARA utility and $\ES_{0.8}$, and a 3\% increase in $\ES_{0.95}$. Specifically, we look at the function $s(L_i, L_j) = \Id_{\{L_i > \Finv_i(0.95)\}} \,\Id_{\{L_j > \Finv_j(0.95)\}}$, where $\Finv_i(\cdot)$ and $\Finv_j(\cdot)$ are the $\P$-quantile functions of $L_i$ and $L_j$ respectively. This bivariate sensitivity measures quantifies the impact a stress has on the probability of joint exceedances with values $\S_{5,10} ^{\Q^*_2} = 0.78$ and $\S_{9,10} ^{\Q^*_2} = 0.81$ indicating that the probabilities of joint exceedances increase more for stress 2.  
This can also be seen in Figure \ref{fig:HARA-ES-contour} which shows the bivariate copulae contours of $(L_5, L_{10})$ (top panels) and $(L_9, L_{10})$ (bottom panels). The left contour plots correspond to the baseline model $\P$ whereas the right panels display the contours under the stress model $\Q^*_2$ (solid lines) together with the baseline contours (reported using partially transparent lines). The red dots are the simulated realisations of the losses $(L_5, L_{10})$ and $(L_9, L_{10})$, respectively (which are the same for the baseline and stressed model). We observe that for both pairs $(L_5, L_{10})$ and $(L_9, L_{10})$ the copula under the stressed model admits larger probabilities of joint large events, which is captured by the bivariate reverse sensitivity measure admitting positive values close to 1.

\section{Concluding Remarks}
We extend the reverse  sensitivity analysis proposed by \cite{Pesenti2019EJOR} which proceeds as follows. Equipped with a baseline model which comprises of input and output random variables and a baseline probability measure, one derives a unique stressed model such that the output (or input) under the stressed model satisfies a prespecified stress and is closest to the baseline distribution. While \cite{Pesenti2019EJOR} consider the Kullback-Leibler divergence to measure the difference between the baseline and stressed models we utilise Wasserstein distance of order two. Compared to \cite{Pesenti2019EJOR} the Wasserstein distance allows for additional and different stresses on the output including the mean and variance, any distortion risk measure including the Value-at-Risk and Expected-Shortfall, and expected utility type constraints, thus making the reverse sensitivity analysis framework suitable for models used in financial and insurance risk management.
We further discuss reverse sensitivity measures which quantify the change the inputs' distribution when moving from the baseline to a stressed model and illustrate our results on a spatial insurance portfolio application. The reverse sensitivity analysis framework (including the results from this work and from \cite{Pesenti2019EJOR} are implemented in the \texttt{R} package \texttt{SWIM} which is available on \texttt{CRAN}.

\appendixtitles{no} 
\appendixstart
\appendix

\section{Proofs}
\begin{proof}[Proof of Theorem \ref{thm:rm-constraints}]
We solve the optimisation on the set of quantile functions $\Minv$ and define the Lagrangian with Lagrange multipliers $\blambda = (\lambda_1, \ldots, \lambda_d) \in \R^d$
\begin{subequations}
\begin{align*}
    L(\Ginv, \blambda) 
    & = \int_0^1 \left(\Ginv(u) - \Finv(u)\right)^2 - 2\,\sum_{k = 1}^d \lambda_k\left( \Ginv(u)  \gamma_k(u)- r_k\right) \; du
    \\
    & =
    \int_0^1 \left(\Ginv(u) - \left(\Finv(u) + \sum_{k = 1}^d \lambda_k \,\gamma_k(u)\right)\right)^2  
    \\
    &    \qquad  
    - 2\,\sum_{k = 1}^d \lambda_k \,\left( \Finv(u)\gamma_k(u) - r_k\right) 
    - \left(\sum_{k = 1}^k \lambda_k  \gamma_k(u) \right)^2\; du\,.
\end{align*}
\end{subequations}
Thus, the optimisation problem \eqref{opt:1} is equivalent to first solving, for fixed $\blambda$, the optimisation problem
\begin{equation}\label{pf:eq:opt}
    \argmin_{\Ginv \in \Minv} L(\Ginv, \blambda)
\end{equation}
and then finding $\blambda$ such that the constraints are fulfilled. For fixed $\blambda$, the solution to \eqref{pf:eq:opt} is equal to the solution to
\begin{equation*}
    \argmin_{\Ginv \in \Minv} \;\int_0^1 \left(\Ginv(u) - \left(\Finv(u) + \sum_{k = 1}^d \lambda_k \,\gamma_k(u)\right)\right)^2\; du\,,
\end{equation*}
which is given by the isotonic projection of $\Finv(u) + \sum_{k = 1}^d \lambda_k \,\gamma_k(u)$ onto the set $\Minv$ and the Lagrange multipliers are such that the constraints are satisfied. Existence of the Lagrange multipliers follows since the set $\M$ is non-empty. Uniqueness follows by convexity of the Wasserstein distance, by convexity of the constraints on the set of quantile functions.
\end{proof}

\begin{proof}[Proof of Proposition \ref{prop:coherent-rm-const}]
For coherent distortion risk measures the corresponding weight function $\gamma$ is non-decreasing. Moreover the optimal quantile function is given by Theorem \ref{thm:rm-constraints} and is of the form $\Ginv_\lambda(u) = \left(\Finv(u) + \lambda \gamma(u)\right)^\uparrow$ for some $\lambda$ such that $\Ginv_\lambda$ fulfils the constraint. The choice
\begin{equation*}
    \lambda^*  = \frac{r - \rho_\gamma(F)}{\int_0^1 \left(\gamma(u)\right)^2\, du} \;\ge 0
\end{equation*}
implies that $\Ginv_{\lambda^*}(u) = \Finv(u) +\lambda^*\gamma(u)$ is a quantile function of the form \eqref{opt:1:sol} that fulfils the constraint. By uniqueness of Theorem \ref{thm:rm-constraints}, $\Ginv_{\lambda^*}$ is indeed the unique solution. 
\end{proof}

\begin{proof}[Proof of Theorem \ref{thm:mult-int-constraints}]
Since both constraints are convex in $\Ginv$ the Lagrangian with parameters $\blambda = (\lambda_1, \ldots, \lambda_d)$ and $\tilde{\blambda}  = (\tilde{\lambda}_1, \ldots, \tilde{\lambda}_{\tilde{d}})\ge 0$ becomes
\begin{subequations}
\begin{align*}
L(\Ginv, \blambda, \tilde{\blambda})
    &=  \int_0^1 \left(\Ginv(u) - \Finv(u)\right)^2 + 2\,\sum_{k = 1}^d \lambda_k\left(  h_k(u) \Ginv(u) - c_k\right) \; du\\
    &  \quad + \,\sum_{k = 1}^{\tilde{d}} \tilde{\lambda}_k\left(\tilde{h}_k(u)\left(\Ginv(u)\right)^2  - \tilde{c}_k\right) \; du\\
    &=  \int_0^1 \tilde{\Lambda}(u) \;\left(\Ginv(u) -\frac{1}{\tilde{\Lambda}(u)}\left( \Finv(u) - \sum_{k = 1}^d \lambda_k  h_k(u)\right)\right)^2\\
    & \quad -
    \frac{1}{\tilde{\Lambda}(u)}\left( \Finv(u) - \sum_{k = 1}^d \lambda_k  h_k(u)\right)^2   
    + \left(\Finv(u)\right)^2
    - 2\sum_{k = 1}^{d} \lambda_k c_k
    - \sum_{k = 1}^{\tilde{d}} \tilde{\lambda}_k \tilde{c}_k\,,
\end{align*}
\end{subequations}
where $\tilde{\Lambda}(u)= 1 +\sum_{k = 1}^{\tilde{d}} \tilde{\lambda}_k\tilde{h}_k(u)$. Since $\tilde{\blambda} \ge 0$ by the KKT-condition, we obtain that $\tilde{\Lambda}(u)\ge0$ for all $u \in [0,1]$. Therefore, for  fixed  $\blambda, \, \tilde{\blambda}$, using an argument similar to the proof of Theorem \ref{thm:rm-constraints}, the solution (as a function of $\blambda, \, \tilde{\blambda}$) is given by the weighted isotonic projection of $\frac{1}{\tilde{\Lambda}(u)}\left( \Finv(u) - \sum_{k = 1}^d \lambda_k  h_k(u)\right)$, with weight function $\tilde{\Lambda}(\cdot)$. 
\end{proof}

\begin{proof}[Proof of Proposition \ref{prop:moment-risk-measures}]
The mean and variance constraint can be rewritten as
\begin{align*}
m^\prime &= \int x \, dG(x) = \int_0^1 \Ginv(u) \, du 
    \qquad \text{and } \quad\\
    \left(\sigma^\prime\right)^2 &= \int \left( x - m^\prime\right)^2\, dG(x) 
    = \int_0^1 \left( \Ginv(u) - m^\prime\right)^2 \, du \,.
\end{align*}
Thus, the Lagrangian with Lagrange multipliers $\blambda = (\lambda_1, \ldots, \lambda_{k+2})$ is, if $\lambda_2 \neq -1 $,
\begin{align*}
    L(\Ginv, \blambda) 
    & = \int_0^1 \left(\Ginv(u) - \Finv(u)\right)^2 \, du
        - 2 \lambda_1 \left(\int_0^1 \Ginv(u) \, du  - m^\prime\right)
        \\
        & \qquad
        +  \lambda_2 \left(\int_0^1 \left( \Ginv(u) - m^\prime\right)^2 \, du  - \left(\sigma^\prime\right)^2\right)\\
    & \quad - 2\,\sum_{k = 1}^{d} \lambda_{k+2}\left(\int_0^1 \Ginv(u)  \gamma_k(u)\, du- r_k\right) \\
    & = (1 + \lambda_2) \int_0^1 \left(\Ginv(u) - \frac{1}{1 + \lambda_2}\left(\Finv(u) + \lambda_1 + \lambda_2 m^\prime
    + \sum_{k = 1}^d \lambda_{k+2} \,\gamma_k(u)\right)\right)^2 \\
    & \quad - \frac{1}{1 + \lambda_2}\, \left(\Finv(u) + \lambda_1 + \lambda_2 m^\prime +\sum_{k = 1}^d \lambda_{k+2}\gamma_k(u) \right)^2
     + \left(\Finv(u)\right)^2\; du\\
    & \quad + 2\lambda_1 m^\prime + \lambda_2\left( \left(m^\prime\right)^2 - \left(\sigma^\prime\right)^2\right)
    + 2 \sum_{k =1}^d \lambda_{k + 2}\,r_k\,.
\end{align*}
For fixed Lagrange multipliers $\blambda$ with $\lambda_2 \neq -1 $, the optimal quantile function is characterised by the isotonic projection and given by (using an analogous argument to the proof of Theorem \ref{thm:rm-constraints})
\begin{subequations}
\begin{align}
    \Ginv^*(u) 
    &= \left(\frac{1}{1 + \lambda_2}\left(\Finv(u) + \lambda_1 + \lambda_2 m^\prime
    + \sum_{k = 1}^{d} \lambda_{k+2} \,\gamma_k(u)\right)\right)^\uparrow \label{eq:iso-mean-var-rm}
    \\
    &= \frac{1}{|1 + \lambda_2|}\left(\sgn(1 + \lambda_2)\left(\Finv(u) + \lambda_1 + \lambda_2 m^\prime
    + \sum_{k = 1}^{d} \lambda_{k+2} \,\gamma_k(u)\right)\right)^\uparrow
    \nonumber
    \\
    &= \frac{1}{|1 + \lambda_2|}\; \Hinv(u)\,,
    \nonumber
\end{align}
\end{subequations}
where we define $\Hinv (u) = \left(\sgn(1 + \lambda_2)\left(\Finv(u) + \lambda_1 + \lambda_2 m^\prime + \sum_{k = 1}^{d} \lambda_{k+2} \,\gamma_k(u)\right)\right)^\uparrow \in \Minv $, and $\sgn(\cdot)$ denotes the sign function. Next we show that $\lambda_2$ cannot be in a neighbourhood of $-1$. It holds that for $\lambda_2 \neq -1$, 
\begin{equation}\label{eqn:mean-var-lambda}
     \int_0^1 \left(\Ginv^*(u)\right)^2 \, du
    = \frac{1}{(1 + \lambda_2)^2}
    \int_0^1 \left(\Hinv(u)\right)^2 \, du\,.
\end{equation}
Since the rhs of \eqref{eqn:mean-var-lambda} is increasing for $|\,\lambda_2 + 1\,| \searrow 0$, there exists a $\ep_0 >0$ such that for all $\ep < \ep_0$ and $\lambda_2 \in ( -1 - \ep, -1+\ep) $, it holds that 
\begin{equation*}
    \frac{1}{(1 + \lambda_2)^2}
    \int_0^1 \left(\Hinv(u)\right)^2 \, du 
   \; > \;
   \left(\sigma^\prime\right)^2 + \left(m^\prime\right)^2\,,
\end{equation*}
which is a contradiction to the optimality of $\Ginv^*$. Thus, $\lambda_2$ is indeed bounded away from $-1$ and the unique solution is given in \eqref{eq:iso-mean-var-rm}.
\end{proof}

\begin{proof}[Proof of Theorem \ref{thm:VaR}]
We split this proof into the two cases $i)$, that is constraint a) and $ii)$, i.e. constraint b). \\
\underline{Case $i)$}: 
For constraint a), i.e. $\VaR_{\alpha}(G) = q$, we first assume that $q \le \VaR_\alpha(F) $ which implies $\Finv(\alpha_F) = q  \le  \Finv(\alpha)$ and thus $\alpha_F \le \alpha$. Therefore, $\Ginv^*(u) = \Finv(u)  + \big(q - \Finv(u)\big)\Id_{\left\{u \in \left(\alpha_F, \alpha\right]\right\}}$ is a quantile function which satisfies the constraint. Next, we show that $G^*$ has a smaller Wasserstein distance to $F$ than any other distribution function satisfying the constraint. For this, let $\Hinv$ be a quantile function satisfying the constraint and $\Hinv(u) \neq \Ginv(u)$ on a measurable set of non-zero measure. Then
\begin{subequations}
\begin{align*}
    W_2(H, F)
    &= \int_0^{\alpha_F} \left(\Hinv(u) - \Finv(u)\right)^2\, du 
    + \int_{\alpha_F}^\alpha \left(\Hinv(u) - \Finv(u)\right)^2\, du 
    +\int_{\alpha}^1 \left(\Hinv(u) - \Finv(u)\right)^2\, du
    \\
    & \ge \int_{\alpha_F}^\alpha \left(\Hinv(u) - \Finv(u)\right)^2\, du\,.
\end{align*}
\end{subequations}
By non-decreasingness of $\Hinv$ and $\Finv$ and by the constraint it holds for all $u \in [\alpha_F, \alpha]$ that $\Hinv(u) \le \Hinv(\alpha)  = q = \Finv(\alpha_F) \le \Finv(u)$. Thus, on the interval $[\alpha_F, \alpha]$, we obtain $\big(\Hinv(u) - \Finv(u)\big)^2 \ge \big(q -  \Finv(u)\big)^2 $ and therefore 
\begin{equation*}
    W_2(H, F)
    \ge \int_{\alpha_F}^\alpha \left(\Hinv(u) - \Finv(u)\right)^2\, du
    \:\ge\: \int_{\alpha_F}^\alpha \left(q - \Finv(u)\right)^2\, du
     = W_2(G^*, F)\,,
\end{equation*}
where at least one inequality is strict since $\Hinv(u) \neq \Ginv(u)$ on a measurable set of non-zero measure. Uniqueness follows by the strict convexity of the Wasserstein distance and since the constraint is convex on the set of quantile functions. 

Second, we assume that $q > \VaR_\alpha(F)$ and show that there does not exist a solution. Assume by contradiction that $\Ginv$ is an optimal quantile function satisfying the constraint. By definition of $\alpha_F$, we have that $q = \Finv(\alpha_F) > \Finv(\alpha)$ and thus $\alpha_F \ge \alpha$.
We apply a similar argument to the first part of the proof using non-decreasingness of $\Ginv$, $\Ginv(\alpha) = q$, and optimality of $\Ginv$, to obtain that $\Ginv$ is constant equal to $q$ on $[\alpha, \alpha_F]$ and equal to $\Finv(u)$ for $u > \alpha_F$. Specifically, it holds that
\begin{equation*}
    \Ginv(u) = \Finv(u) + (q - \Finv(u)) \Id_{\{u \in (\alpha, \alpha_F]\}}\,,
    \, \quad \text{for all} \quad u > \alpha\,.
\end{equation*}
Moreover, since the optimal quantile function minimises the Wasserstein distance to $F$, it holds that, for all $\ep >0$, $\Ginv$ satisfies
\begin{equation*}
    \Ginv(u) = \Finv(u)\,,\quad \text{for all} \quad u \le \alpha - \ep\,.
\end{equation*}
Thus, we can define for all $\ep \in (0, \alpha)$ the family of quantile functions 
\begin{equation*}
    \Hinv_\ep(u) = \Finv(u) + (q - \Finv(u)) \Id_{\{u \in (\alpha - \ep, \alpha_F]\}}\,,
\end{equation*}
which satisfies $W_2(H_{\ep_1}, F) < W_2(H_{\ep_2}, F) $ for all $0 \le \ep_1 < \ep_2$, and $\Hinv_{\ep}(\alpha) = q$ for all $\ep >0$. However, $\lim_{\ep \searrow 0} \Hinv_\ep(\alpha) = \Finv(\alpha) < q$ and thus the quantile function $\lim_{\ep \searrow 0} \Hinv_\ep(u)$ does not fulfil the constraint. Hence, we obtain a contradiction to the optimality of $\Ginv$.

\underline{Case $ii)$}: 
First, we assume that $q \ge \VaR_\alpha^+ (F)$ which implies that $\Finv(\alpha_F) = q \ge \Finv^+(\alpha) \ge \Finv(\alpha)$ and thus $\alpha_F \ge \alpha$. Therefore $\Ginv^*(u) = \Finv(u)  + \big(q - \Finv(u)\big)\Id_{\left\{u \in \left(\alpha, \alpha_F\right]\right\}}$ is a quantile function. Moreover, $\Ginv^*$ satisfies the constraint since by right-continuity of $\Ginv^*$, we have that
\begin{equation*}
    \Ginv^{*\, +}(\alpha)
    = \lim_{\ep \searrow 0} \Ginv^{*\, +}(\alpha + \ep)
    = q\,.
\end{equation*}
The proof that $\Ginv^*$ has the smallest Wasserstein distance to $F$ compared to any other distribution function satisfying the constraint is analogous to the one in case $i)$.

For the case when $q > \VaR_\alpha^+ (F)$, the argument of non-existence of the solution follows using similar arguments as those in case $i)$.
\end{proof}

\begin{proof}[Proof of Theorem \ref{thm:exp-utility-risk-measure}]
By concavity of the utility function, the constraint is convex and can be written as $- \int_0^1 u\left(\Ginv(v)\right)\, dv + c \le 0$. 
Thus, we can define the Lagrangian with $\lambda_1 \ge0$ and $(\lambda_2, \ldots, \lambda_{d+1}) \in \R^d$ by
\begin{subequations}
\begin{align*}
    L(\Ginv, \lambda)
    &= 
    \frac12 \int_0^1 \left(\Ginv(v) - \Finv(v)\right)^2 
    - \lambda_1 \left( u\big(\Ginv(v)\big) \,  - c\right)
    - \sum_{k = 1}^{d} \lambda_{k+1} \left(\Ginv(v) \gamma_k(v) - r_k\right)\,dv
    \\
    &= 
    \int_0^1  T\left(\Ginv(v) \right)
    - \Ginv(v)\left(\Finv(v) + \sum_{k = 1}^{d} \lambda_{k+1} \gamma_k(v)\right) \, 
    \\
    & \qquad
    + \tfrac{1}{2}\left(\Finv(v)\right)^2
    + \lambda_1 \,c + \sum_{k = 1}^{d} \lambda_{k+1} \, r_k
    \;dv\,,
\end{align*}
\end{subequations}
where  $T(x) = \frac12 x^2 -\lambda_1 u(x)$. Therefore, for fixed $\lambda_1, \ldots, \lambda_{d+1}$, we apply Theorem 3.1 by \cite{Barlow1972JASA} and obtain the unique optimal quantile function (as a function of $\lambda_1, \ldots, \lambda_{d+1}$), that is $\Ginv^*(v) = \breve{\nu}_{\lambda_1}\left(\big(\Finv(u)  + \sum_{k = 1}^{d} \lambda_{k+1} \gamma_k(v)\big)^
\uparrow\right)$, where $\breve{\nu}_{\lambda_1}$ is the left-inverse of $ \nu_{\lambda_1}(x) = x  - \lambda_1 \,u^\prime(x)$. 

Next, we show that if $d= 0$, the utility constraint is binding, that is $\lambda_1 >0$. For this, assume by contradiction that the $\lambda_! = 0$, then the optimal quantile function becomes $\Ginv^*(u) = \breve{\nu}_0 \big(\Finv(u) \big)$. Since $\nu_0(x) = x$, we obtain that $\Ginv^*(u) = \Finv(u)$. $\Finv$, however, does not fulfil the constraint, which is a contradiction to the optimality of $\Ginv^*$. 
\end{proof}

\begin{proof}[Proof of Proposition \ref{prop:properties-rev-sens}]
We prove the properties one-by-one:
\begin{enumerate}[label = $\roman*)$]
    \item 
We first define for a random variable $Z$ with $\P$-distribution $F_Z$ the random variable $U_Z := F_Z(Z)$. Then, $U_Z$ and $Z$ are comonotonic and $U_Z$ has a uniform distribution under $\P$. Next, recall that for any random variables $Y_1, Y_2$ it holds that \cite{Ruschendorf1983Metrika}
\begin{equation} \label{eq:como-exp}
\E \left[Y_1\, F_{Y_2}^{-1}\left(1 - U_{Y_1}\right)\right]\le
\E \left[Y_1\, Y_2\right]
\le
\E \left[Y_1\, F_{Y_2}^{-1}\left(U_{Y_1}\right)\right].
\end{equation}
where $F_{Y_2}^{-1}\left(U_{Y_1}\right)$ is the random variable that is comonotonic to $Y_1$ and has the same $\P$-distribution as $Y_2$. Similarly, $F_{Y_2}^{-1}\left(1 - U_{Y_1}\right)$ is the random variable that is counter-monotonic to $Y_1$ and has the same $\P$-distribution as $Y_2$. The left (right) inequality in \eqref{eq:como-exp} become equality if and only if the random variables $Y_1$ and $Y_2$ are counter-comonotonic (comonotonic). 

Thus, we can rewrite the maximum in the normalising constant of the reverse sensitivity measure as follows
\begin{equation*}
    \max\limits_{\Q \in \mQ}\,\E^{\Q} \left[s(X)\right] 
    = \max\limits_{Z \stackrel{\P}{=} \frac{d\Q^*}{d\P}}\E \left[s(X)\, Z\right]\,
     = \E \left[s(X)\, F_{\frac{d\Q^*}{d\P}}^{-1}\left(U_{s(X)}\right)\right]\,
\end{equation*}
and the minimum in the normalising constant is
\begin{equation*}
    \min\limits_{\Q \in \mQ}\E^{\Q} \left[s(X)\right] 
    = \min\limits_{Z \stackrel{\P}{=} \frac{d\Q^*}{d\P}}\E \left[s(X)\, Z\right]\,
     = \E \left[s(X)\, F_{\frac{d\Q^*}{d\P}}^{-1}\left(1 - U_{s(X)}\right)\right]\,.
\end{equation*}
The reverse sensitivity for the case $\E^{\Q^*}[s(X_i)] \ge \E[s(X_i)]$ then becomes
\begin{equation*}
    S^{\Q^*}_i
    = \frac{\E [s(X_i)\frac{d\Q^*}{d\P} ] - \E[s(X_i)]}{ \E \left[s(X)\, F_{\frac{d\Q^*}{d\P}}^{-1}\left(U_{s(X)}\right)\right] - \E[s(X_i)]}\,,
\end{equation*}
which satisfies $0 \le S^{\Q^*}_i\le 1$ using again \eqref{eq:como-exp}. For the case $\E^{\Q^*}[s(X_i)] \le \E[s(X_i)]$, it holds that
\begin{equation*}
    S^{\Q^*}_i
    = - \frac{\E [s(X_i)\frac{d\Q^*}{d\P} ] - \E[s(X_i)]}{ \E \left[s(X)\, F_{\frac{d\Q^*}{d\P}}^{-1}\left(1 - U_{s(X)}\right)\right] - \E[s(X_i)]}\,,
\end{equation*}
which satisfies $-1 \le S^{\Q^*}_i\le 0$.

\item Assume that $s(X_i)$ and $\frac{d\Q^*}{d\P}$ are independent under $\P$, then 
\begin{equation*}
    \E [s(X_i)\tfrac{d\Q^*}{d\P} ] 
    =
    \E \left[s(X_i)\right]\E\left[\tfrac{d\Q^*}{d\P} \right] 
    =
    \E [s(X_i)]\,,
\end{equation*}
and the reverse sensitivity measure is indeed zero.

\item From property $i)$ we observe that $s(X_i)$ and $\frac{d\Q^*}{d\P}$ are comonotonic, if and only if, $S^{\Q^*}_i = 1$ since in this case the right inequality in Equation \eqref{eq:como-exp} becomes equality. 

\item From property $i)$ we observe that $s(X_i)$ and $\frac{d\Q^*}{d\P}$ are counter-comonotonic, if and only if, then $S^{\Q^*}_i = 1$ as in this case left inequality in Equation \eqref{eq:como-exp} becomes equality.  \\[0.5em]
\end{enumerate}

The proof that the joint reverse sensitivity $S_{i,j}^{\Q^*}$ also fulfils the above properties follows using analogous arguments and replacing $s(X_i)$ with $s(X_i, X_j)$.
\end{proof}

\section*{Acknowledgments}
SP would like to thank Judy Mao for her help in implementing the numerical examples. \\
SP gratefully acknowledges the support of the Connaught Fund, the Canadian Statistical Sciences Institute (CANSSI), and the Natural Sciences and Engineering Research Council of Canada (NSERC) with funding reference numbers DGECR-2020-00333 and RGPIN-2020-04289.

\begin{adjustwidth}{-\extralength}{0cm}

\reftitle{References}



\bibliography{references}


\end{adjustwidth}
\end{document}

%% file: preamble.tex
\usepackage{dsfont}
\usepackage{amssymb, amsmath}
\usepackage{enumitem}
\usepackage{multicol, multirow}
\usepackage{booktabs}
\newcommand{\ep}{{\varepsilon}}
\newcommand{\Id}{{\mathds{1}}}

\newcommand{\Cor}{{\textrm{Cor}}}

\newcommand{\E}{{\mathbb{E}}}
\newcommand{\N}{{\mathbb{N}}}
\renewcommand{\P}{{\mathbb{P}}}
\newcommand{\Q}{{\mathbb{Q}}}
\newcommand{\R}{{\mathbb{R}}}
\newcommand{\Lp}{{\mathbb{L}^2}}

\newcommand{\x}{{\boldsymbol{x}}}
\newcommand{\z}{{\boldsymbol{z}}}

\newcommand{\X}{{\boldsymbol{X}}}

\renewcommand{\z}{{\boldsymbol{z}}}

\newcommand{\blambda}{{\boldsymbol{\lambda}}}

\renewcommand{\S}{{\mathcal{S}}}
\newcommand{\mQ}{{\mathcal{Q}}}
\newcommand{\Minv}{{\Breve{\mathcal{M}}}}
\newcommand{\Ginv}{{\Breve{G}}}
\renewcommand{\Finv}{{\Breve{F}}}
\newcommand{\Hinv}{{\Breve{H}}}
\newcommand{\M}{{\mathcal{M}}}

\newcommand{\VaR}{{\textrm{VaR}}}
\newcommand{\RVaR}{{\textrm{RVaR}}}
\newcommand{\ES}{{\textrm{ES}}}

\DeclareMathOperator{\sgn}{{sgn}}
\DeclareMathOperator*{\argmin}{{\mathrm{arg}\min}}

\usepackage{todonotes}


%% file: main.bbl
\begin{thebibliography}{999}

\bibitem[Saltelli \em{et~al.}(2008)Saltelli, Ratto, Andres, Campolongo,
  Cariboni, Gatelli, Saisana, and Tarantola]{Saltelli2008book}
Saltelli, A.; Ratto, M.; Andres, T.; Campolongo, F.; Cariboni, J.; Gatelli, D.;
  Saisana, M.; Tarantola, S.
\newblock {\em Global sensitivity analysis: the primer}; John Wiley \& Sons,
  2008.

\bibitem[Borgonovo and Plischke(2016)]{Borgonovo2016EJOR}
Borgonovo, E.; Plischke, E.
\newblock Sensitivity analysis: a review of recent advances.
\newblock {\em European Journal of Operational Research} {\bf 2016}, {\em
  248},~869--887.

\bibitem[Tsanakas and Millossovich(2016)]{Tsanakas2016RA}
Tsanakas, A.; Millossovich, P.
\newblock Sensitivity analysis using risk measures.
\newblock {\em Risk Analysis} {\bf 2016}, {\em 36},~30--48.

\bibitem[Maume-Deschamps and Niang(2018)]{Maume2018SPL}
Maume-Deschamps, V.; Niang, I.
\newblock Estimation of quantile oriented sensitivity indices.
\newblock {\em Statistics \& Probability Letters} {\bf 2018}, {\em
  134},~122--127.

\bibitem[Asimit \em{et~al.}(2019)Asimit, Peng, Wang, and Yu]{Asimit2019MF}
Asimit, V.; Peng, L.; Wang, R.; Yu, A.
\newblock An efficient approach to quantile capital allocation and sensitivity
  analysis.
\newblock {\em Mathematical Finance} {\bf 2019}, {\em 29},~1131--1156.

\bibitem[Fissler and Pesenti(2022)]{Fissler2022sensitivity}
Fissler, T.; Pesenti, S.M.
\newblock Sensitivity Measures Based on Scoring Functions.
\newblock {\em arXiv preprint arXiv:2203.00460} {\bf 2022}.

\bibitem[Borgonovo \em{et~al.}(2016)Borgonovo, Hazen, and
  Plischke]{Borgonovo2016RA}
Borgonovo, E.; Hazen, G.B.; Plischke, E.
\newblock A common rationale for global sensitivity measures and their
  estimation.
\newblock {\em Risk Analysis} {\bf 2016}, {\em 36},~1871--1895.

\bibitem[Borgonovo(2007)]{Borgonovo2007RESS}
Borgonovo, E.
\newblock A new uncertainty importance measure.
\newblock {\em Reliability Engineering \& System Safety} {\bf 2007}, {\em
  92},~771--784.

\bibitem[Rahman(2016)]{Rahman2016UQ}
Rahman, S.
\newblock The f-sensitivity index.
\newblock {\em SIAM/ASA Journal on Uncertainty Quantification} {\bf 2016}, {\em
  4},~130--162.

\bibitem[Gamboa \em{et~al.}(2018)Gamboa, Klein, and Lagnoux]{Gamboa2018SIAMUQ}
Gamboa, F.; Klein, T.; Lagnoux, A.
\newblock Sensitivity Analysis Based on {C}ram\'er--von {M}ises Distance.
\newblock {\em SIAM/ASA Journal on Uncertainty Quantification} {\bf 2018}, {\em
  6},~522--548.

\bibitem[Plischke and Borgonovo(2019)]{Plischke2019EJOR}
Plischke, E.; Borgonovo, E.
\newblock Copula theory and probabilistic sensitivity analysis: Is there a
  connection?
\newblock {\em European Journal of Operational Research} {\bf 2019}, {\em
  277},~1046--1059.

\bibitem[Gamboa \em{et~al.}(2020)Gamboa, Gremaud, Klein, and
  Lagnoux]{Gamboa2020ARXIV}
Gamboa, F.; Gremaud, P.; Klein, T.; Lagnoux, A.
\newblock Global Sensitivity Analysis: a new generation of mighty estimators
  based on rank statistics.
\newblock {\em arXiv preprint arXiv:2003.01772} {\bf 2020}.

\bibitem[Pesenti \em{et~al.}(2021)Pesenti, Millossovich, and
  Tsanakas]{Pesenti2021RA}
Pesenti, S.M.; Millossovich, P.; Tsanakas, A.
\newblock Cascade sensitivity measures.
\newblock {\em Risk Analysis} {\bf 2021}, {\em 31},~2392--2414.

\bibitem[Denuit \em{et~al.}(2006)Denuit, Dhaene, Goovaerts, and
  Kaas]{Denuit2006book}
Denuit, M.; Dhaene, J.; Goovaerts, M.; Kaas, R.
\newblock {\em Actuarial theory for dependent risks: measures, orders and
  models}; John Wiley \& Sons,  2006.

\bibitem[Pesenti \em{et~al.}(2019)Pesenti, Millossovich, and
  Tsanakas]{Pesenti2019EJOR}
Pesenti, S.M.; Millossovich, P.; Tsanakas, A.
\newblock Reverse sensitivity testing: What does it take to break the model?
\newblock {\em European Journal of Operational Research} {\bf 2019}, {\em
  274},~654--670.

\bibitem[Cambou and Filipovi{\'c}(2017)]{Cambou2017MF}
Cambou, M.; Filipovi{\'c}, D.
\newblock Model uncertainty and scenario aggregation.
\newblock {\em Mathematical Finance} {\bf 2017}, {\em 27},~534--567.

\bibitem[Pesenti \em{et~al.}(2021)Pesenti, Bettini, Millossovich, and
  Tsanakas]{Pesenti2020AAS}
Pesenti, S.M.; Bettini, A.; Millossovich, P.; Tsanakas, A.
\newblock Scenario Weights for Importance Measurement ({SWIM})--an {R} package
  for sensitivity analysis.
\newblock {\em Annals of Actuarial Science} {\bf 2021}, {\em 15},~458--483.

\bibitem[Makam \em{et~al.}(2021)Makam, Millossovich, and
  Tsanakas]{Makam2021IME}
Makam, V.D.; Millossovich, P.; Tsanakas, A.
\newblock Sensitivity analysis with $\chi$2-divergences.
\newblock {\em Insurance: Mathematics and Economics} {\bf 2021}, {\em
  100},~372--383.

\bibitem[Kruse \em{et~al.}(2019)Kruse, Schneider, and Schweizer]{Kruse2019OR}
Kruse, T.; Schneider, J.C.; Schweizer, N.
\newblock The joint impact of F-divergences and reference models on the
  contents of uncertainty sets.
\newblock {\em Operations Research} {\bf 2019}, {\em 67},~428--435.

\bibitem[Bernard \em{et~al.}(2020)Bernard, Pesenti, and
  Vanduffel]{Bernard2020WP}
Bernard, C.; Pesenti, S.M.; Vanduffel, S.
\newblock Robust distortion risk measures.
\newblock {\em Available at SSRN} {\bf 2020}.

\bibitem[Blanchet and Murthy(2019)]{Blanchet2019MOR}
Blanchet, J.; Murthy, K.
\newblock Quantifying distributional model risk via optimal transport.
\newblock {\em Mathematics of Operations Research} {\bf 2019}, {\em
  44},~565--600.

\bibitem[Moosm\"ueller \em{et~al.}(2020)Moosm\"ueller, Dietrich, and
  Kevrekidis]{Moosmueller2021SIAMUQ}
Moosm\"ueller, C.; Dietrich, F.; Kevrekidis, I.G.
\newblock A geometric approach to the transport of discontinuous densities.
\newblock {\em SIAM/ASA Journal on Uncertainty Quantification} {\bf 2020}, {\em
  8},~1012--1035.

\bibitem[Fort \em{et~al.}(2021)Fort, Klein, and Lagnoux]{Fort2021global}
Fort, J.C.; Klein, T.; Lagnoux, A.
\newblock Global sensitivity analysis and Wasserstein spaces.
\newblock {\em SIAM/ASA Journal on Uncertainty Quantification} {\bf 2021}, {\em
  9},~880--921.

\bibitem[Villani(2008)]{Villani2008book}
Villani, C.
\newblock {\em Optimal transport: Old and new}; Vol. 338, Springer Science \&
  Business Media,  2008.

\bibitem[Dall'Aglio(1956)]{dall1956SNS}
Dall'Aglio, G.
\newblock Sugli estremi dei momenti delle funzioni di ripartizione doppia.
\newblock {\em Annali della Scuola Normale Superiore di Pisa-Classe di Scienze}
  {\bf 1956}, {\em 10},~35--74.

\bibitem[Barlow \em{et~al.}(1972)Barlow, Bartholomew, Bremner, and
  Brunk]{Barlow1972book}
Barlow, R.E.; Bartholomew, D.; Bremner, J.M.; Brunk, H.D.
\newblock {\em Statistical inference under order restrictions: the theory and
  application of isotonic regression}; Wiley,  1972.

\bibitem[De~Leeuw \em{et~al.}(2010)De~Leeuw, Hornik, and Mair]{De2010JSS}
De~Leeuw, J.; Hornik, K.; Mair, P.
\newblock Isotone optimization in R: pool-adjacent-violators algorithm (PAVA)
  and active set methods.
\newblock {\em Journal of Statistical Software} {\bf 2010}, {\em 32},~1--24.

\bibitem[Acerbi and Tasche(2002)]{Acerbi2002JBF}
Acerbi, C.; Tasche, D.
\newblock On the coherence of {E}xpected {S}hortfall.
\newblock {\em Journal of Banking \& Finance} {\bf 2002}, {\em 26},~1487--1503.

\bibitem[Artzner \em{et~al.}(1999)Artzner, Delbaen, Eber, and
  Heath]{Artzner1999MF}
Artzner, P.; Delbaen, F.; Eber, J.M.; Heath, D.
\newblock Coherent measures of risk.
\newblock {\em Mathematical Finance} {\bf 1999}, {\em 9},~203--228.

\bibitem[Kusuoka(2001)]{Kusuoka2001AME}
Kusuoka, S.
\newblock On law invariant coherent risk measures. In {\em Advances in
  Mathematical Economics}; Springer,  2001; pp. 83--95.

\bibitem[Cont \em{et~al.}(2010)Cont, Deguest, and Scandolo]{Cont2010QF}
Cont, R.; Deguest, R.; Scandolo, G.
\newblock Robustness and sensitivity analysis of risk measurement procedures.
\newblock {\em Quantitative Finance} {\bf 2010}, {\em 10},~593--606.

\bibitem[Sysoev and Burdakov(2019)]{Sysoev2019SPAV}
Sysoev, O.; Burdakov, O.
\newblock A smoothed monotonic regression via L2 regularization.
\newblock {\em Knowledge and Information Systems} {\bf 2019}, {\em
  59},~197--218.

\bibitem[Hall and Huang(2001)]{Hall2001AS}
Hall, P.; Huang, L.S.
\newblock Nonparametric kernel regression subject to monotonicity constraints.
\newblock {\em The Annals of Statistics} {\bf 2001}, {\em 29},~624--647.

\bibitem[Meyer(2008)]{Meyer2008AAS}
Meyer, M.C.
\newblock Inference using shape-restricted regression splines.
\newblock {\em The Annals of Applied Statistics} {\bf 2008}, {\em
  2},~1013--1033.

\bibitem[Cuestaalbertos \em{et~al.}(1993)Cuestaalbertos, Ruschendorf, and
  Tuerodiaz]{Cuestaalbertos1993JMA}
Cuestaalbertos, J.A.; Ruschendorf, L.; Tuerodiaz, A.
\newblock Optimal coupling of multivariate distributions and stochastic
  processes.
\newblock {\em Journal of Multivariate Analysis} {\bf 1993}, {\em
  46},~335--361.

\bibitem[Borgonovo \em{et~al.}(2021)Borgonovo, Hazen, Jose, and
  Plischke]{Borgonovo2021EJOR}
Borgonovo, E.; Hazen, G.B.; Jose, V.R.R.; Plischke, E.
\newblock Probabilistic sensitivity measures as information value.
\newblock {\em European Journal of Operational Research} {\bf 2021}, {\em
  289},~595--610.

\bibitem[Barlow and Brunk(1972)]{Barlow1972JASA}
Barlow, R.E.; Brunk, H.D.
\newblock The isotonic regression problem and its dual.
\newblock {\em Journal of the American Statistical Association} {\bf 1972},
  {\em 67},~140--147.

\bibitem[R{\"u}schendorf(1983)]{Ruschendorf1983Metrika}
R{\"u}schendorf, L.
\newblock Solution of a statistical optimization problem by rearrangement
  methods.
\newblock {\em Metrika} {\bf 1983}, {\em 30},~55--61.

\end{thebibliography}
